\documentclass[useAMS,usegraphicx,usenatbib,a4paper]{mn2e}
\usepackage[]{amsmath}
\usepackage{aas_macros}
\usepackage{color}
\usepackage[dvips, bookmarks, pdfborder={0 0 0.1}]{hyperref}
\usepackage{xspace}
\usepackage{multirow, multicol}

\newcommand{\D}{\mathrm{d}}

\newcommand{\msunh}{\mathrm{M}_\odot h^{-1}}
\graphicspath{{figures/}}

\newcommand{\myplot}[1]{\includegraphics[width=0.5\textwidth]{#1}}
\newcommand{\myplottwo}[2]{\myplot{#1}\myplot{#2}}

\newcommand{\mytab}{\begin{table}[htb]}
\newcommand{\myfig}{\begin{figure}[htbp]}

\newcommand{\mybibstyle}{mymn}

\newcommand{\erf}{\mathrm{erf}}

\newcommand{\hbtp}{\textsc{hbt$+$}\xspace}
\newcommand{\hbt}{\textsc{hbt}\xspace}
\newcommand{\rockstar}{\textsc{rockstar}\xspace}
\newcommand{\subfind}{\textsc{subfind}\xspace}
\newcommand{\surv}{\textsc{surv}\xspace}
\newcommand{\dtree}{\textsc{dtree}\xspace}
\newcommand{\subtree}{\textsc{subfind}$+$\textsc{dtree}\xspace}
\newcommand{\LCDM}{$\Lambda$CDM\xspace}
\newcommand{\openmp}{\textsc{openmp}\xspace}
\newcommand{\mpi}{\textsc{mpi}\xspace}

\begin{document}
\title[\hbtp]{{\sc hbt$+$}: an improved code for finding subhalos and
  building merger trees in cosmological simulations}  
\author[J. Han et al.] 
{Jiaxin Han,$^{1,2}$\thanks{jiaxin.han@ipmu.jp} Shaun Cole,$^2$ Carlos S. Frenk,$^2$ Alejandro Benitez-Llambay,$^2$ John Helly$^2$\\ 
$^1$Kavli IPMU (WPI), UTIAS, The University of Tokyo, Kashiwa, Chiba 277-8583, Japan\\
$^2$Institute of Computational Cosmology, Department of Physics, University of Durham, South Road, Durham, DH1 3LE, UK\\} 
\maketitle

\begin{abstract} 
Dark matter subhalos are the remnants of (incomplete) halo mergers. Identifying them and establishing their evolutionary links in the
form of  merger trees is one of the most important applications of cosmological simulations. The Hierachical Bound-Tracing (\hbt)
code identifies halos as they form and tracks their evolution as they merge, simultaneously detecting subhalos and building their merger trees. Here we present a new implementation of this approach, \hbtp, that is much faster, more user friendly, and more physically complete than the original code. Applying \hbtp to cosmological simulations we show that both the subhalo mass function and the peak-mass function are well fit by similar double-Schechter functions.The ratio between the two is highest at the high mass end, reflecting the resilience of massive subhalos that experience substantial dynamical friction but limited tidal stripping. The radial distribution of the most massive subhalos is more concentrated than the universal radial distribution of lower mass subhalos. Subhalo finders that work in configuration space tend to underestimate the masses of massive subhalos, an effect that is stronger in the host centre. This may explain, at least in part, the excess of massive subhalos in galaxy cluster centres inferred from recent lensing observations. We demonstrate that the peak-mass function is a powerful diagnostic of merger tree defects, and the merger trees constructed using \hbtp do not suffer from the missing or switched links that tend to afflict merger trees constructed from more conventional halo finders. We make the \hbtp code publicly available.
\end{abstract}
  
\section{Introduction}
The process of cosmic structure formation, as revealed by numerical
simulations, can be largely summarized by the growth of dark matter halos and their
interactions. In a universe in which the dark matter is cold (CDM),
small halos merge hierarchically to form bigger halos, a process that
is often described by a halo merger tree. After a merger, the remnants 
of the progenitors are not erased immediately. Instead, they survive
inside the descendant halo as subhalos~\citep{Moore98, skid, Klypin99,
  Moore99}. A complete list of halos and subhalos, together with their
merger history, has become a standard data product of a simulation,
whose calculation requires a (sub)halo finder and a merger tree
builder.

Finding isolated dark matter halos is relatively straightforward
once a definition of halo is adopted. For example, a
Friends-of-Friends (\textsc{FoF}) halo finder~\citep{FoF} works by
connecting particles located within a linking length of each other to
find clustered particles above a certain density threshold. A
spherical overdensity halo finder~\citep[e.g.,][]{LC94} works by
growing a radius around a density peak until the average density
inside the sphere matches a predefined value. By contrast, finding
subhalos are more complicated. Generally speaking, the process of
finding a subhalo consists of two steps: 1) collecting a list of
candidate particles to build a ``source'' subhalo; 2) pruning the
source to remove unbound particles until a self-bound subhalo
remains.

Depending on the way the source is defined, subhalo finders can be
broadly categorized into three types: configuration space finders that
examine the spatial clustering of particles~\citep[e.g.,][]{subfind,
  ahf, skid}; phase space finders that consider  clustering in both
spatial and velocity space~\citep[e.g.,][]{velociraptor,
  rockstar, hsf}; and tracking finders that build the source from past
progenitors~\citep{surv,HBT}. It has been shown that configuration
space finders suffer from a ``blending'' problem, the difficulty of
resolving subhalos embedded in the inner high density region of the
host halo due to spatial overlap ~\citep{stuart, mad, HBT}. In the case of major mergers, this problem is further
manifest as a random switching of the masses of the merging halos or
of the presumed halo centre: once the two protagonists of the merger
overlap substantially, the partitioning of mass between them can be
arbitrary and inconsistent from snapshot to snapshot. Even phase space
finders have difficulty dealing with this situation~\citep{major}.
These problems in identifying the main descendant of a merger propagate
into the merger tree, giving rise to incorrect or missing
links~\citep{HBT, suss, suss_avila,DHalo}.

One way to solve these problems is by exploiting prior knowledge about
the history of the subhalo particles. A tracking finder such as the Hierarchical Bound-Tracing~\citep[\hbt][]{HBT} achieves this by taking the list of particles in the
progenitor as the source of the subhalo. This approach relies on
the fact that a subhalo can be defined as the self-bound remnant of
its progenitor halo after a merger. Since \hbt does not rely on
spatial or phase space clustering to build the source, it is naturally
immune to the blending and mass or centre-switching  problems. 

In this work, we present a new implementation of the \hbt algorithm,
\hbtp. \hbtp is written in \textsc{C++} from scratch, and improves
upon \hbt in many respects including modularity, usability,
performance, support for distributed architecture, applicability to
hydrodynamical simulations, and richness in output subhalo properties.
The default output format is \textsc{hdf5} which can be easily
manipulated in scripting languages such as \textsc{python}. Besides
the technical improvements, the most significant change in the
physical prescription is that \hbtp can handle the merger of subhalos
due to dynamical friction. It is known that \hbt catalogues include
subhalos that are located at nearly identical positions in phase
space~\citep{HBT, major}. Although it may be desirable to track these
overlapping objects separately for certain applications, their
separate identities are not supported by the resolution of the
simulation. In \hbtp we introduce a prescription to detect and merge
these overlapping pairs. As we will show, this mostly affects the
population of surviving subhalos at the high mass end and in the very
centre of the host halo.

The downside of the tracking approach is that it does not work on a
single snapshot of the simulation. Instead, a sequence of snapshots
has to be provided. This problem can be overcome by combining a
tracking finder with the simulation code and carrying out halo finding
and tree building on-the-fly. The optimized \hbtp would be a good
candidate for such developments.

We apply \hbtp to cosmological and zoomed simulations to test the
performance of the code. As an illustration we consider the
distribution of massive subhalos. Recent lensing observations
suggest an excess of massive subhalos in galaxy clusters. 
Using a combined weak and strong lensing analysis, 
\citet{Jauzac16} and \citet{Schwinn} compared the distribution of
massive subhalos inferred in Abell 2744 with those in the
Millennium-XXL~\citep{Angulo_12} simulation. They could not find any
halo in Millennium-XXL that hosts as many massive subhalos as are observed in Abell 2744. \citet{HSTFF} carried out a strong lensing analysis of the subhalo distribution in the inner region of several galaxy clusters in the Hubble Frontier Field, and compared their results with the Illustris~\citep{Illustris} hydrodynamical simulations. They find relatively good agreement in the subhalo mass function, but the observed radial distribution of subhalos is more concentrated than that in simulations. These comparisons are based on subhalo catalogues constructed from \subfind~\citep{subfind}, a subhalo finder in configuration space.
Besides selection effects and extreme number statistics, these authors
interpret the discrepancy as due to overly efficient dynamical
friction and tidal stripping in the simulation. Interestingly, it has
been argued that configuration space finders such as \subfind
significantly underestimate the subhalo mass function at the high mass
end~\citep{BJ16}. This conclusion is mostly based on comparing
\subfind results with those from \rockstar\citep{rockstar} and
\surv\citep{surv, G08, G10}. Unfortunately, these studies did include 
a comparison of the radial distribution of subhalos. In this work we
compare both the mass and radial distributions of \hbtp and \subfind
subhalos in detail, by applying both finders to the same set of
simulations.

Our analysis also eliminates some systematic uncertainties in the
\citet{BJ16} comparison. One such uncertainty is in the definition of
subhalo mass. While \rockstar and \surv define the mass of a subhalo
to include the contribution from its sub-subhalos, \subfind follows an
exclusive mass definition. In addition, the \subfind results used in
that comparison were inferred from a fitting formula derived from a
different simulation with a different convention for defining the host
halo properties. In this work, we will make a direct comparison
between \subfind and \hbtp by applying them to the same set of
simulations. Since both \subfind and \hbtp adopt an exclusive mass
definition for subhalos, this allows a fair comparison. 

We confirm the conclusion of \citet{BJ16} that \subfind underestimates
the high mass end of the mass function. We find that this deficiency
is mostly caused by the difficulty \subfind has in resolving massive
subhalos near the host centre. This means that the excess of massive
subhalos in cluster centres may be attributed to systematics in the
subhalo catalogues in the simulations, rather than posing a challenge
to current \LCDM cosmological simulations. This ``indigestion'' of
massive subhalos leads to a hardening in the subhalo mass function
at the high mass end, which explains the flatter slope found by
\citet{BJ16}. However, at much lower subhalo masses, we find a slope
for the subhalo mass function of $-0.95$, consistent with previous
studies \cite{Springel08}

This paper is organized as follows. In Section~\ref{sec:algorithm} we
explain the technical details of the \hbtp algorithm, with 
emphasis on the improvements over its predecessor, \hbt. In
Section~\ref{sec:tests} we apply \hbtp to simulations to test its
performance, with special attention to the distribution of massive
subhalos. We summarize and conclude in Section~\ref{sec:summary}.

\section{Algorithm}\label{sec:algorithm}
In this paper, we make an explicit distinction between a halo and a
subhalo. A halo is defined as an \emph{isolated} virialized object,
while a subhalo is a substructure embedded inside a halo. As an input
to \hbtp, an existing halo catalogue containing the list of particles
in each isolated halo at each snapshot must be provided. This
halo-finding step can be done with any halo finder of the user's
preference, and is independent of the subhalo finding step which is
the main function of \hbtp.  

\begin{figure}
\myplot{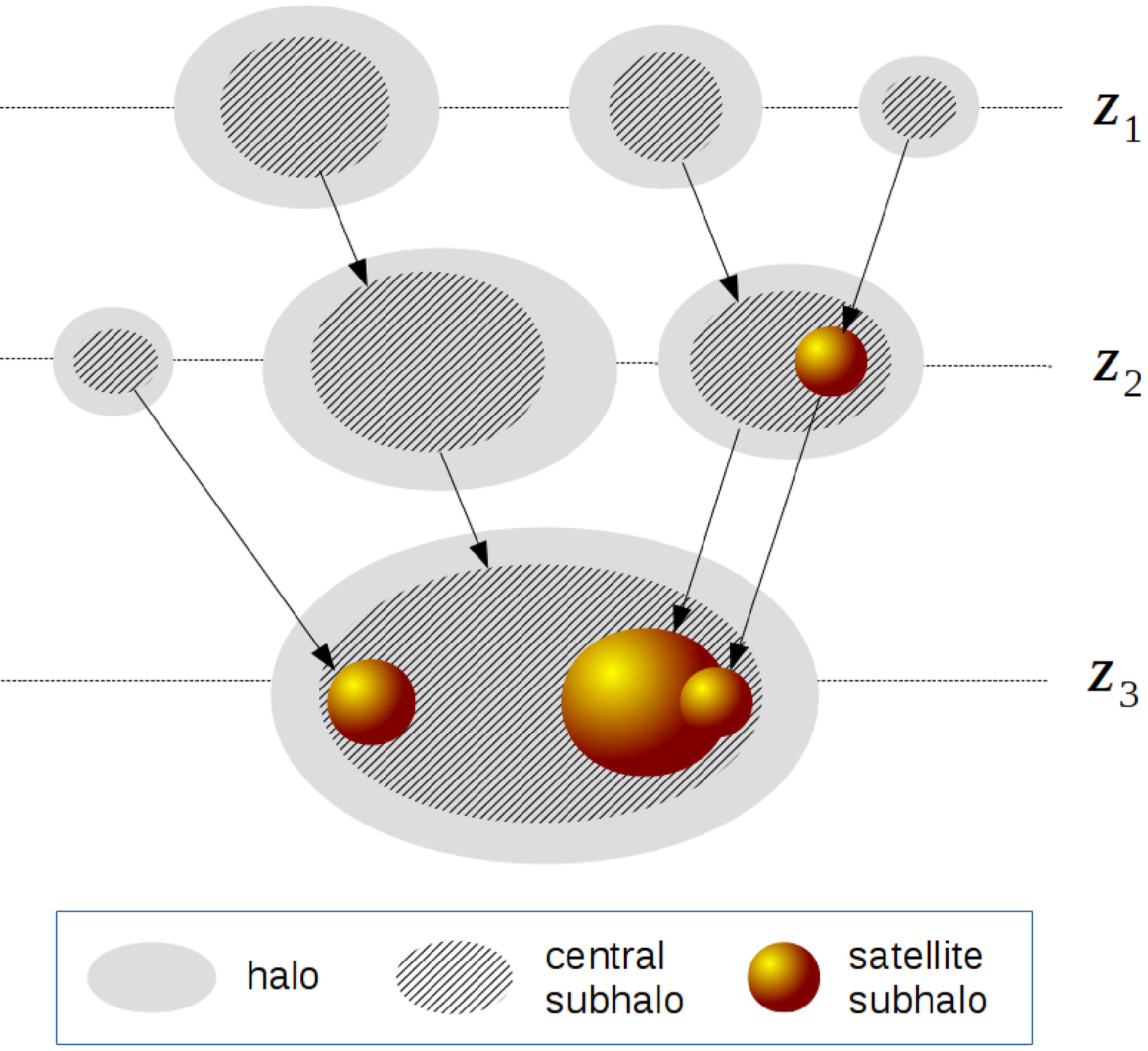}
\caption{Illustration of a merger tree and the algorithm to find subhalos through tracking. Each horizontal line represents a snapshot labelled by its redshift, with $z_1>z_2>z_3$. As halos merge, satellite subhalos are created as remnants of the progenitors, and can be identified by tracking the progenitor particles to subsequent snapshots and removing unbound particles. For each bound halo, a central subhalo is always defined as the one containing the majority of its bound particles, and is linked to the main progenitor.}\label{fig:tree}
\end{figure}

The overall algorithm of \hbtp is the same as that of \hbt, and can be
summarized in Fig.~\ref{fig:tree}. Starting from the earliest snapshot
at redshift $z_1$ of a simulation, each halo is screened to identify
bound particles and eliminate unbound ones. The particle list of each bound
halo is then passed to the next snapshot $z_2$ to identify a
descendant halo. When multiple progenitors are linked to the same
descendant, a main progenitor is determined (according to mass and
dynamical consistency, see Section~\ref{sec:tracking}), while the
others are unbinded to create satellite subhalos. The main progenitor
is then updated to include the particles of the current host halo
(excluding satellites), and unbinded to create a central subhalo. The
distinction between centrals and satellites is used to reflect that a
central subhalo is accreting mass while a satellite is subject to mass stripping. 
The particle lists of these central and satellite subhalos
are further propagated to the next snapshot $z_3$, and the iteration
continues until all the snapshots have been processed.  

The detailed implementation of this algorithm has been improved in
many aspects in \hbtp, and is describe below. 

\subsection{A simple and intuitive merger tree format}
Conventional approaches to subhalo finding and merger tree building works by first finding subhalos at each snapshot, then linking them across snapshots. As a result, each subhalo is regarded as a different object and a tree is represented by the links between subhalos at different snapshots. \hbt also follows this scheme in tree building, by assigning separate subhalo IDs to subhalos in different snapshots, and recording the progenitor ID for each subhalo, as shown in Fig.~\ref{fig:tree}. Such a representation is also commonly found in popular databases (e.g., the Millennium database)\footnote{\url{http://gavo.mpa-garching.mpg.de/Millennium/}, \url{http://galaxy-catalogue.dur.ac.uk:8080/Millennium/}} and in a community proposed merger tree format~\citep{TreeFormat}, where additional auxilliary links are further provided to facilitate tree walking.

\begin{figure}
\includegraphics[width=0.5\textwidth,trim={0 7cm 0 0},clip]{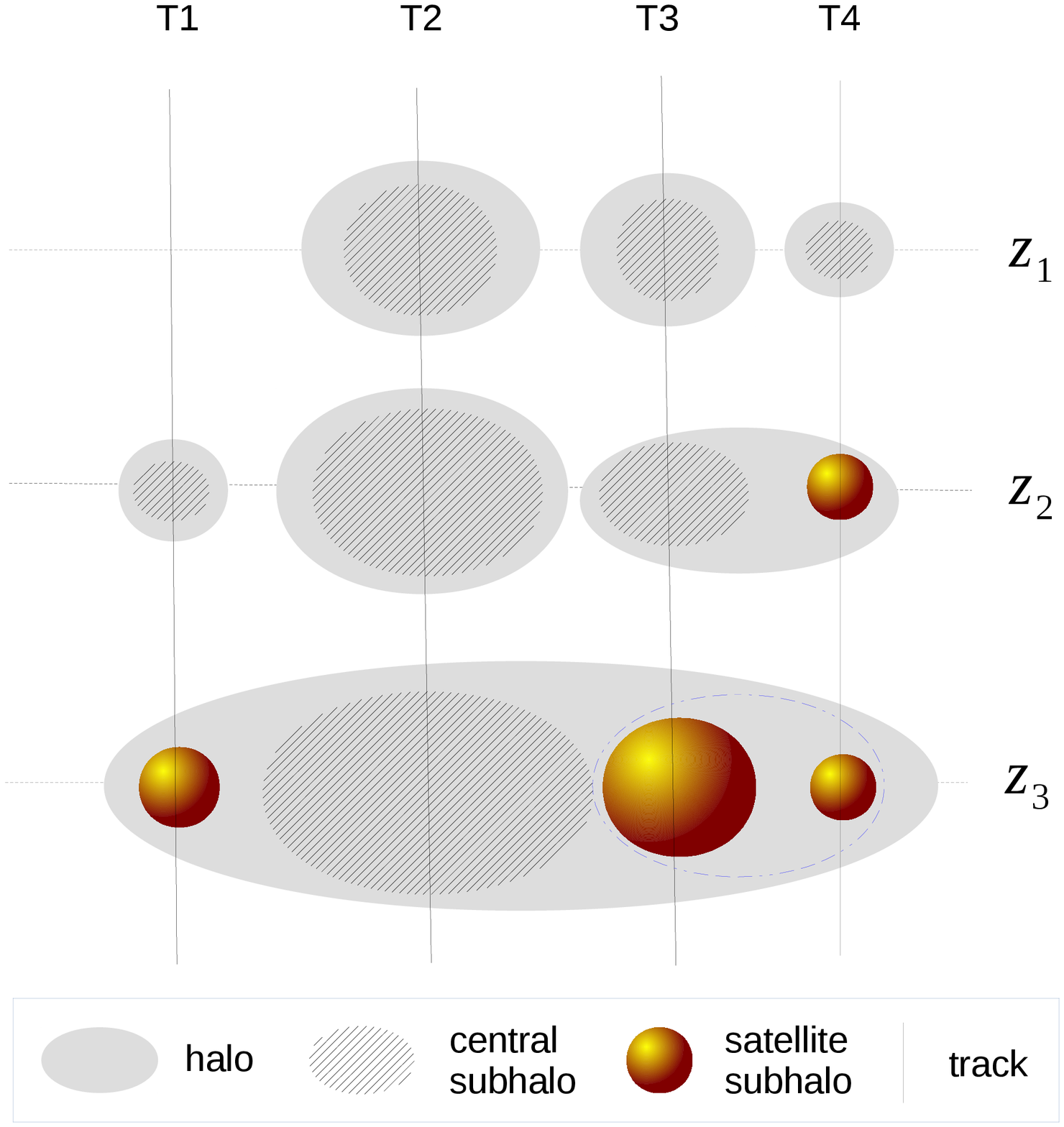}
 \caption{The merger tree in Fig.~\ref{fig:tree} represented as a list
   of tracks ($T1$, $T2$, $T3$ and $T4$) grouped by different host
   halos at each snapshot. The dash-dotted ellipse at $z_3$
   marks a subhalo group.} 
\label{fig:track}
\end{figure}

In \hbtp, we switch to an alternative representation of subhalos and
trees by organizing them in terms of tracks, that are native to the
tracking algorithm. Each track is the entire evolution history of a
subhalo, while a subhalo is a snapshot of a track. This is equivalent
to treating a subhalo as a Lagrangian object, which is labelled by a
single Lagrangian ID throughout time. Thus a merger tree can be
completely specified by a list of tracks associated with halos at each
snapshot (e.g., by recording a host halo ID for each track).
Fig.~\ref{fig:track} shows such an example. Properties of each subhalo
at each snapshot can still be added as a local property of the track
at different times. This approach essentially flattens the merger tree
into a table, which is much more convenient and flexible to store.
Such a ``track table'' is more convenient to query and sample as well.
For example, one can directly obtain the progenitor or descendant of a
subhalo at any other snapshot by searching for the given track ID,
without having to walk the tree snapshot by snapshot. One can also
freely remove arbitrarily selected snapshots from the catalogue
without having to rebuild the merger tree. As in \hbt, the
merging hierarchy is propogated to subsequent snapshots to record
subhalo groups, so that some subhalos can be satellites of another
subhalo.
  
Once a track ID is created, it persists through all following
snapshots. When a subhalo's mass drops below the mass resolution of
the simulation, we use the most bound particle to represent the track.
This can be useful for galaxy formation models that place ``orphan
galaxies'' on top of these most bound particles. It can also be used
to identify subhalo mergers by identifying the host subhalo of this
most bound particle when the subhalo disappears. 


\subsection{Tracking}~\label{sec:tracking}
\textbf{Host finding:}
For each subhalo, its host halo at the next snapshot is simply
determined to be the host halo of its most-bound particle.\footnote{In
  some rare cases, a subhalo does not find any host halo but remains
  bound. This could happen when the host halo is occasionally missed
  by the halo finder (e.g., \textsc{FoF}) near the resolution limit. We keep
  these types of objects as field subhalos and assign the background
  universe as a special host halo for them.} We have checked that such
a tracking is robust enough compared with tracking multiple most-bound
particles. This is much cleaner than the original \hbt treatment that
splits the progenitor particles into different hosts, which mostly
introduces short-lived noisy tracks (splitter tracks) into the
catalogue. 

\textbf{Main progenitor determination:}
Inside each host halo, the main progenitor is typically selected to be
the most-massive one. However, when other progenitors have masses
close to the most massive one, such a choice becomes less justified.
In this case, we further compare the kinetic energy of the progenitors
with respect to the bulk motion of the host halo. Out of all the
progenitors whose mass exceeds $2/3$ of the most-massive progenitor mass,
the one that has the smallest specific kinetic energy is chosen to be
the main progenitor. As we further justify in
Equation~\ref{eq:KineticDist} in Section~\ref{sec:stripping}, this
choice yields the highest total binding energy when all the halo
particles are accreted by the main progenitor. \footnote{After
  unbinding, the subhalos are sorted in mass one more time to ensure
  that the most massive bound subhalo is assigned as the central
  subhalo.}


\textbf{Source subhalo update:} An important step for robust tracking is
selecting a set of particles--a source subhalo-- for each subhalo
that are passed to the next snapshot for unbinding
(Section~\ref{sec:stripping}). As is shown in \citet{HBT}, the definition of the source subhalo
has to be precise enough so as to avoid too many unbound particles,
while at the same time it has to be conservative enough to allow for
reaccretion of previously stripped particles. In \hbt this is achieved
by adaptively chosing a progenitor at some previous snapshot according
to the current mass of the subhalo. In this work, we do this in a more
flexible way by updating the source continuously. After each
unbinding step, source particles are sorted according to binding
energy. Less bound particles are excluded, to leave a source
subhalo with at most $3N_{\rm bound}$ particles, where $N_{\rm bound}$
is the number of bound particles. This updated source is then
passed to the next snapshot for unbinding.

\subsection{Stripping}\label{sec:stripping}
The stripping of mass from subhalos is determined by unbinding.

\textbf{Reference frame.}
Unbinding is the process of removing particles whose kinetic energy exceeds their potential energy. To calculate the kinetic energy, a reference frame must be defined, which we choose to be the one that minimizes the total kinetic energy of the subhalo particles. Since 
\begin{align}
 K_{\rm tot}&=\frac{1}{2}\sum \limits_{i=1}^{N} (\vec{v}_i-\vec{v}_{\rm c})^2\\
 &=\frac{N}{2} [(\langle\vec{v}\rangle-\vec{v}_{\rm c})^2+ (\langle\vec{v}^2\rangle-\langle\vec{v}
 \rangle^2)],\label{eq:KineticDist}
\end{align} minimizing the kinetic energy is equivalent to minimizing the distance between the centre velocity and the average velocity vectors, $(\langle\vec{v}\rangle-\vec{v}_{\rm c})^2$. When Hubble flow is considered, the distance becomes $(\langle\vec{v}\rangle-\vec{v}_{\rm c}+H\langle\vec{r}-\vec{r}_{\rm c}\rangle)^2$. A natural choice is thus the centre of mass frame, centred at $\vec{r}_{\rm c}=\langle\vec{r}\rangle$ with bulk velocity $\vec{v}_{\rm c}=\langle\vec{v}\rangle$. Once unbound particles are removed, we update the reference frame and calculate the binding energies using the gravitational potential from the remaining particles, and unbind again. This process continues until the bound mass converges.

\textbf{Fast unbinding:}
The calculation of potential energy during the unbinding iteration is expensive even with a tree code. The majority of the computation time of \hbt is spent on unbinding. We introduce two optimizations to speed up this process. 

The first optimization is to apply a differential potential update. During every step of the unbinding iteration, the change in potential is due to the removal of unbound particles. When the number of removed particles between two iterations becomes smaller than that of the remaining particles, the potential energy can be efficiently obtained by applying a correction to the potential in the previous iteration, that is, by subtracting the contribution from the removed particles.

For the purpose of unbinding, a very accurate potential energy is not required. Thus we further optimize this step by calculating the potential using a small sample of randomly selected subhalo particles. As we show analytically in Appendix~\ref{sec:unbind_noise}, the bound density profile of a subhalo can be recovered to percent level accuracy or better over the entire radial range when the mass distribution is sampled with only $1000$ particles. 
With this algorithm, the potential calculation for all the $N$ particles in the subhalo becomes an $O\left(N\right)$ operation, compared to the $O\left(N\log(N)\right)$ complexity of a tree-code.


Because the potential energy is less accurate in the centre when calculated with a sampled mass distribution, it becomes difficult to select the most bound particle which is a commonly adopted reference frame of a subhalo. To overcome this problem, we further calculate an ``inner binding energy'' adopting only the potential from the $1000$ most bound particles, and select the most-bound particle thereafter.

We also tried unbinding using a potential estimate that assumes spherical symmetry by binning the mass distribution radially, which can also speed up the calculation significantly. However, such an unbinding tends to fail when spherical symmetry is not a good approximation, such as near pericentric passage where the tidal shear is strong.


\begin{figure}
 \myplot{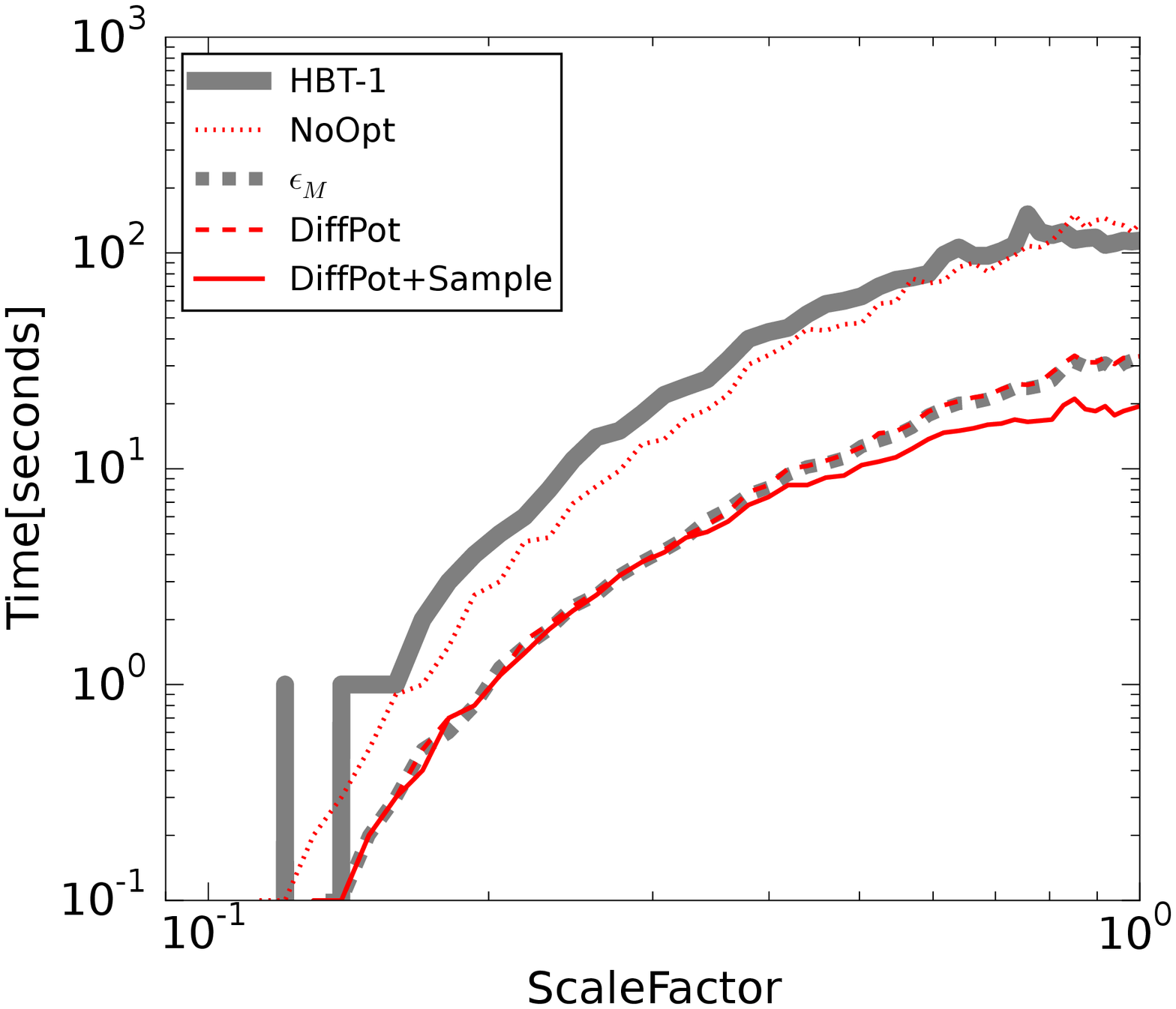}
 \caption{The runtime of \hbtp on a test simulation. Different curves
   show the performance with different levels of optimizations:
   \textit{NoOpt:} \hbtp with no optimization; \textit{DiffPot:} \hbtp
   with only differential potential update; \textit{DiffPot$+$Sample:}
   \hbtp with both differential potential update and sampled potential
   estimate. For comparison, the thick grey line (\textsc{HBT-1}) shows
   the timing of \hbt, and the dashed grey line ($\epsilon_M$) shows
   the timing of \hbtp adopting the same optimization as \hbt, which
   is by relaxing the mass convergence criterion.}\label{fig:timing} 
\end{figure}

In Fig.~\ref{fig:timing} we show the performance improvement achieved by the various optimizations. We use a test simulation of $270^3$ particles with a boxsize of $62.5\, {\rm Mpc}h^{-1}$, run in the same cosmology as that of the Millennium simulation~\citep{Millennium}. The tests are done on a single computational node of the COSMA machine in Durham with $12$ cores. The performance is improved significantly with the differential potential update optimization, and further when the sampled potential estimate is also used. To compare against the performance of \hbt, we also run \hbtp using the same level of optimization as \hbt, which is by terminating the unbinding iteration when the bound mass, $M_{i}$, at iteration $i$ converges with $M_{i+1}/M_{i}>\epsilon_M$ where $\epsilon_M=0.995$ is the mass precision. The performance difference between \hbt and \hbtp adopting this common unbinding optimization can be mostly attributed to a change in the central-determination step in Section~\ref{sec:tracking}. In \hbt, the centrals are selected by comparing the bound mass of the progenitors at the current snapshot, which are obtained by one extra unbinding step. By contrast, in \hbtp, this is done by comparing the progenitor mass at the previous snapshot, together with their current kinetic energies in the host halo frame, avoiding the extra unbinding. Because the $\epsilon_M$ optimization introduces similar improvement as the differential potential update optimization, we no longer rely on the $\epsilon_M$ optimization in \hbtp, although this parameter is still available. Overall, the performance is already increased by a factor of $\sim 6$ for this small test simulation, and we expect even higher improvements for larger simulations, given the change in complexity from $O\left(N\log(N)\right)$ to $O\left(N\right)$.

\textbf{Recursive unbinding}
As in \hbt, the unbinding is done recursively, by unbinding the deepest nested subhalos first and then feeding the stripped particles to their host subhalos for unbinding. This ensures that the particles in each subhalo do not include any bound particles contained in its sub-subhalos, leading to an exclusive mass definition. The subhalo nesting hierarchy is propagated from the merging hiearchy of their progenitor halos.

\subsection{Merging}
\begin{figure}
 \myplot{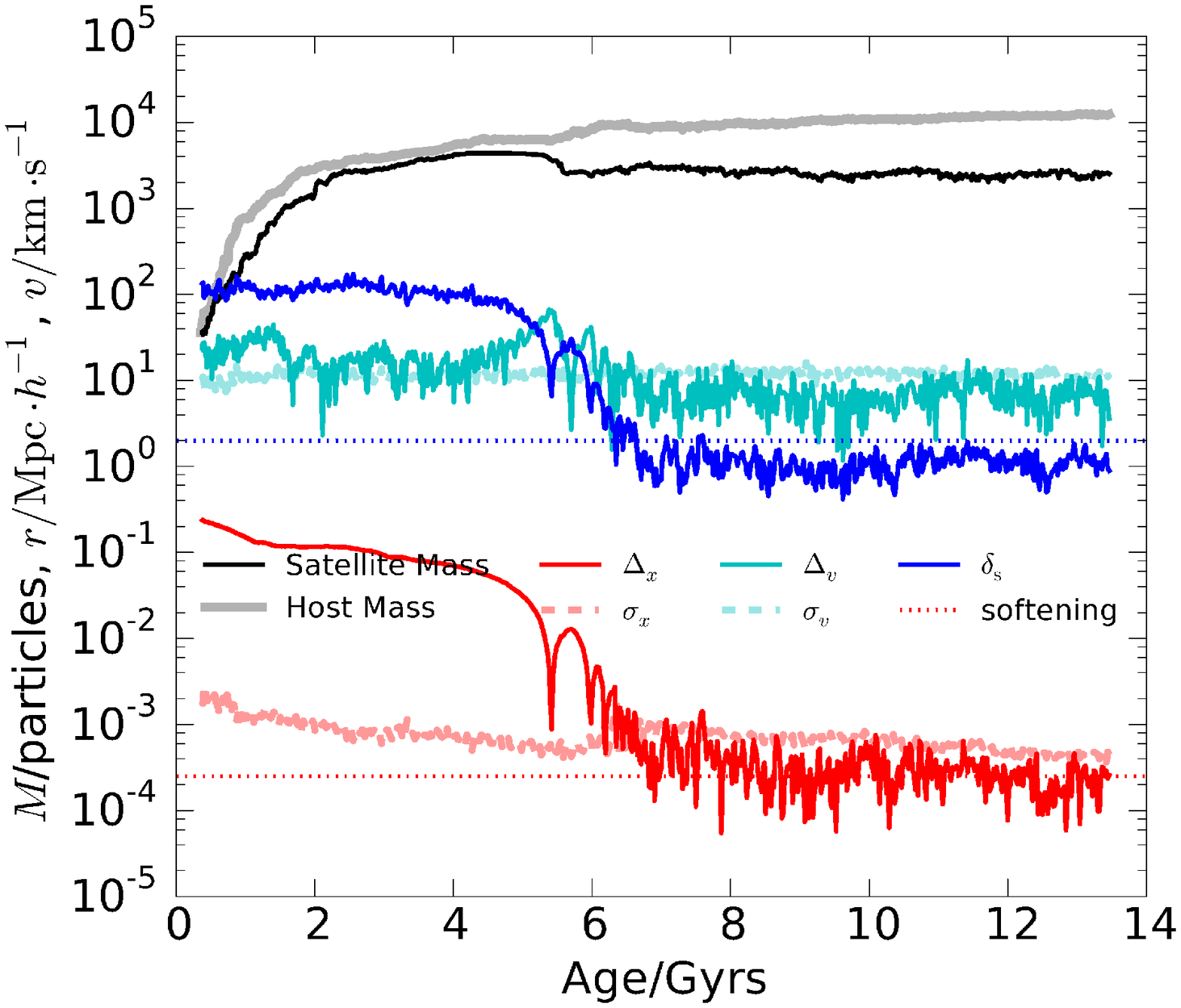}
 \caption{A resolved merger of two subhalos from the Aquarius
   simulation of a Milky Way sized halo with a particle mass of
   $2.9\times 10^5\, \msunh$. We show the evolution of mass, separation
   ($\Delta_x$) and relative velocity ($\Delta_v$)  of the
   two objects. The spatial and velocity dispersions ($\sigma_x$ and
   $\sigma_v$) of the 20 most-bound particles of the host subhalo are
   also shown; these measure the spatial and velocity resolution at
   the centre of the halo. The merger of the two objects can be
   identified with $\delta_{\rm s}<2$, shown as a blue dotted line.}

\label{fig:sink_example} 
\end{figure}

Due to dynamical friction and heating, subhalos could lose their
orbital energy and eventually sink to the centre of their host. When
the trajectories of two subhalos overlap and evolve together without
being separated thereafter, the satellite is trapped at the host
centre, and the two subhalos can be defined as having merged. After a
merger, tidal stripping ceases to take effect due to the concentric
configuration, and the bound mass of the trapped subhalo remains
constant until it is heated up by mass accretion or stripped by other
halos. We show an example in Fig.~\ref{fig:sink_example}. We identify
trapped mergers by comparing the spatial and velocity separations,
$\Delta_x$ and $\Delta_v$, of the two subhalos with the resolution at
the centre of the host subhalo. To estimate the resolution, we use the
spatial and velocity dispersion of the 20 most bound particles of the
host subhalo, $\sigma_x$ and $\sigma_v$. Let   
\begin{equation}
 \delta_{\rm s}=\frac{\Delta_x}{\sigma_x}+\frac{\Delta_v}{\sigma_v}\label{eq:delta}.
\end{equation}
When $\delta_{\rm s}<2$, the two objects are regarded as merged
giving the current numerical resolution. We use the position and
velocity of the most bound particle of each subhalo to measure
$\Delta_x$ and $\Delta_v$, so this merger criterion can be interpreted
as when the most-bound particles of the two objects cannot be
separated in phase space. By default, we merge the trapped subhalo
with its host once $\delta_{\rm s}<2$ and only track the most-bound
particle thereafter. In Appendix~\ref{sec:sink} we provide more
information about the distribution of these trapped subhalos for the
case when we do not implement this merging criterion. 

\subsection{Parallelization}
\hbtp comes in two flavours of parallelization: one pure \openmp version to be used on a shared memory machine, and one \mpi/\openmp hybrid version to be used on distributed servers which can also be run in pure \mpi mode.

For the \openmp version, the parallelization is automatically determined by the symmetry of the workload. When the most massive halo in a snapshot exceeds $10\%$ the total mass of all halos, the parallelization is done inside each halo, by calculating the binding energy of invidividual particles in parallel. Otherwise, the parallelization is done by processing different halos in parallel.

In the current \mpi version, the workload is decomposed by dividing the simulation box into spatial grids which are assigned to different computational nodes. Halos and subhalos are then distributed to the grids according to their spatial coordinates. On each node, the computation is then performed in the same way as the \openmp version. Because the particles loaded from each halo typically only contain the particle IDs, they must be matched to the particles in the snapshot files to obtain their coordinates and other properties. This is done in parallel by first distributing the snapshots to different nodes according to the particle IDs. The particles in each halo are then split and passed to different nodes according to their IDs. For all the halo particles received on each node, we then sort both the halo particles and the snapshot particles. The sorted halo particles are then matched to the snapshot particles with a batched binary search, which successively narrows down the search range of each particle by the search result of the previous particles. The same is done to query subhalo particles.

\subsection{Support for hydrodynamical simulations}
The same tracking and unbinding procedure can be applied to
hydrodynamical simulations no matter how many types of particles exist
in the halo catalogue, although additional routines are needed to
handle the creation of particles due to star formation and the
destruction of particles due to accretion by black holes. Inside each
subhalo, the binding energy of each particle is calculated under the
gravitional potential of all types of particles in the subhalo
adopting a common reference frame. By default, we do not include the
thermal energy of gas particles in the binding energy calculation in
order to reflect the instantaneous dynamical state of the system. The
effect of thermal energy on the system will be automatically revealed
by the instantaneous dynamical state of the system in subsequent
snapshots, once the thermal energy is converted to kinetic energy.
Technically, however, the code can be configured to output both the
binding energy and thermal energy of each particle, so that one can
always switch to an alternative binding energy definition including
the thermal energy in postprocessing. 

\section{Tests and application}\label{sec:tests}
Previous works have already revealed a few features of \hbt, as summarized below:
\begin{itemize}
\item The subhalos found by \hbt have more extended density profiles compared to the truncated outer density profile typical in configuration space finders, leading to a larger mass estimate in \hbt~\citep{HBT}. This difference can be much more significant for massive subhalos.
\item \hbt easily overcomes the blending problem of subhalos, and
  successfully recovers subhalos even when a subhalo is deeply
  embedded in the halo centre~\citep{HBT, suss}.
 \item \hbt maintains consistent link between subhalo progenitor and
   descendant by construction, and is free from the mass or
   centre-switching problem
   in merger tree construction~\citep{suss, major}. 
\end{itemize}

However, the mass difference, the blending and mass or centre-switching
problems are demonstrated only through case studies of individual
objects or through idealized simulations of a single pair of objects.
In this section, we aim to investigate these differences statistically
using cosmological simulations. In particular, we will compare the
distribution of massive subhalos found by \hbtp and \subfind, in order
to understand the systematics in the distribution of these objects,
and to shed light on the observed excess of massive subhalos in
clusters. In fact, the mass difference also can be understood as a
blending problem that obscures the outskirts of a subhalo, while the
mass or centre-switching is created by partial or total obscuration of the subhalo in
the merger tree. It is thus natural to expect that all three issues
are significant for massive subhalos.

\subsection{Simulations}
We make use of two simulations in this section. The first is the Millennium-II simulation~\citep{Mill2} in the \LCDM cosmology with $\Omega_{\rm m}=0.25$ and $\sigma_8=0.9$. It resolves $2160^3$ particles in a cubic box of $100~{\rm Mpc} h^{-1}$ on each side, with a particle mass of $6.89 \times 10^6\, \msunh$. The other simulation is a zoomed in simulation of a Milky Way sized halo from the Aquarius simulation set~\citep{Aquarius} with the same cosmology as Millennium-II. The Aquarius set consists of several Milky Way sized halos each simulated with a series of resolutions. We mainly use the first halo simulated at the second highest resolution level, which we call halo AqA2 hereafter. It has a particle mass of $10^{4}\,\msunh$ in the high resolution region, corresponding to $\sim 10^8$ particles resolved in the main halo. 

The Millennium-II simulation provides a large sample of halos to study
the average distribution of subhalos in host halos of different mass.
Most importantly, it allows us to study the distribution of massive
subhalos statistically, which is impossible with a single host halo
due to the rarity of massive subhalos. We will study three aspects of
the subhalo population: the final subhalo mass, the peak subhalo mass
(i.e., the maximum mass attained by a subhalo over its entire history), 
and the location inside the host halo. As we
will show, combining the final and peak subhalo mass function allows
us to assess the quality of the merger tree statistically, which is
further demonstrated in a side-by-side comparison of the merger trees
of the AqA2 halo.

\subsection{The subhalo mass function}

\begin{figure}
 \myplot{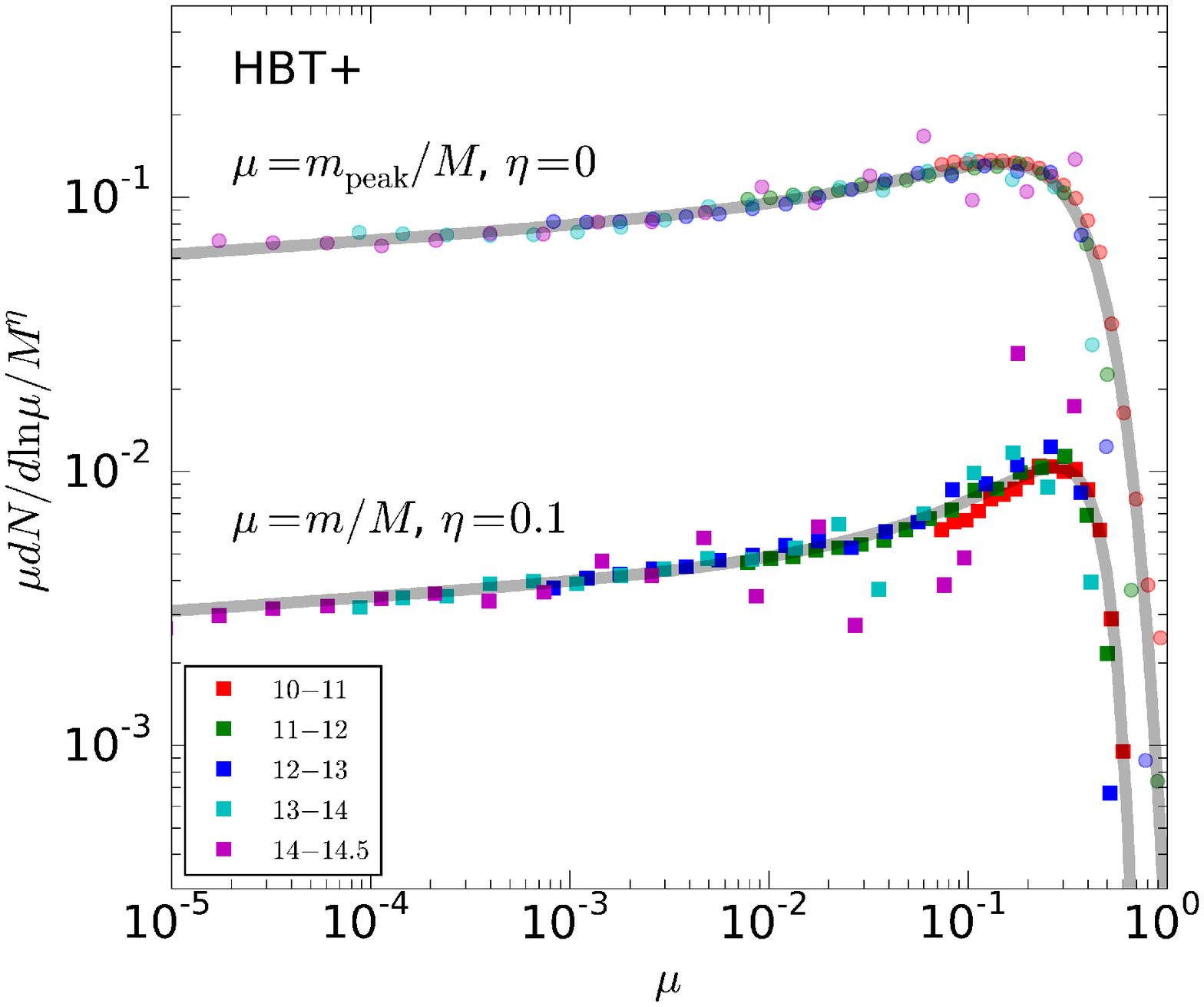}
 \caption{The peak and final subhalo mass functions in Millennium-II
   halos, normalized by $M^\eta$ where $M$ is the host halo mass.
   $\eta$ and the mass variable $\mu$ are specified in the figure for
   the peak and final mass functions respectively. Data points and
   lines of different colour represent different host halo mass bins,
   as listed in the legends in terms of $\log(M/{\msunh})$. The light
   thick lines are the fits of Eqn~\eqref{eq:mf}, with best-fit
   parameters listed in Table~\ref{table:par} (200Crit rows). The
   results for other virial definitions are qualitatively
   similar.}\label{fig:Mill2MF} 
\end{figure}

It is well known that the subhalo mass function follows a simple
power-law behaviour at the low mass end, $\frac{\D N}{\D ln m} \propto
m^{-\alpha}$, with $\alpha\approx0.9$~\citep[e.g.,][]{Gao}. It has
also been shown that the slope of $\alpha$ is conserved between the
unevolved and evolved subhalo mass function~\citep{SubGen}. In
Fig.~\ref{fig:Mill2MF} we show the subhalo mass functions from the
Millennium-II simulation. Both the evolved mass function and the
unevolved mass functions are shown, which we call the final and peak
subhalo mass functions respectively. These functions are computed as
follows. For each host halo, we identify all the branches that are
currently located within its virial radius according to the position
of the most bound particle of each branch. After that, the evolved
mass function is defined as the distribution of the final subhalo mass
of these branches, and the unevolved mass function is defined as the
distribution of their peak bound masses. Both surviving and disrupted
branches contribute to the peak-mass function.

\begin{table*}
\caption{Fits to the subhalo mass functions of the form Equation~\eqref{eq:mf}. We show the results for three different definitions of the host halo mass $M$, corresponding to spherical overdensities of 200 times the critical density (200Crit, our default choice in this work), 200 times the mean mass density (200Mean) of the universe and that given by the spherical collapse model (Virial). The abundances are computed inside the radius of the corresponding spherical overdensity. $m$ and $m_{\rm peak}$ are the final and peak mass of the subhalo. 
}
\label{table:par}
\begin{center}
\begin{tabular}{|c|c|ccccccc|}
\hline
\hline Host Halo Definition & $\mu$ & $a_1$ & $\alpha_1$ & $a_2$ & $\alpha_2$ & $b$ & $\beta$ & $\eta$\\
\hline \multirow{4}{*}{200Crit} & $m/M$       & 0.0055 & 0.95 & 0.017 & 0.24 & 24 & 4.2 & 0.1 \\ 
\cline{2-9}                  & $m_{\rm peak}/M$ & 0.11 & 0.95 & 0.20 & 0.30 & 7.6 & 2.1 & 0  \\
\hline \multirow{4}{*}{Virial} & $m/M$ & 0.0072 & 0.95 & 0.017 & 0.26 & 54 & 4.6 & 0.1 \\ 
\cline{2-9}                  & $m_{\rm peak}/M$ & 0.11 & 0.95 & 0.32 & 0.08 & 8.9 & 1.9 & 0 \\
\hline \multirow{4}{*}{200Mean} & $m/M$ & 0.0090 & 0.95 & 0.055 & 0.16 & 36 & 3.2 & 0.1 \\ 
\cline{2-9}                  & $m_{\rm peak}/M$ & 0.11 & 0.95 & 0.64 & -0.20 & 11 & 1.8 & 0 \\
\hline
\end{tabular}			
\end{center}
\end{table*}

With the large sample of halos, we are able to well resolve the high mass end of the mass function. Both distributions are well fitted by a double Schechter function of the form
\begin{align}
 &f(\mu)\equiv\frac{\D N}{\D \ln \mu}\nonumber \\
 &= \left(\frac{M}{10^{10} \msunh}\right)^{\eta} \left( a_1 \mu^{-\alpha_1}+ a_2 \mu^{-\alpha_2} \right) \exp \left( - b \mu^{\beta} \right),\label{eq:mf}
\end{align}
where $\mu$ is the ratio of the final (peak) mass of the subhalo to
the host halo mass. By default, we adopt the virial definition
corresponding to a spherical overdensity of 200 times the critical
density of the universe. However, we list the best-fit parameters for
three common virial definitions in Table~\ref{table:par}. The first
power-law component in Equation~\eqref{eq:mf} describes the low mass
end behaviour of the mass function, while the second component is
necessary to fit the shoulder at $\mu>0.1$. Note that fitting the low
mass end slope can be tricky depending on the weights given to the
data points, the available mass range that is trusted to have
converged, and the functional form adopted to describe the high mass
end behaviour. Thus we refrain from fitting this slope. Instead, we
fix $\alpha_1=0.95$ according to the result of \citet{SubGen} using
much higher resolution data. As shown in Fig.~\ref{fig:Mill2MF}, such
a choice is well supported by the
data.

\begin{figure}
 \myplot{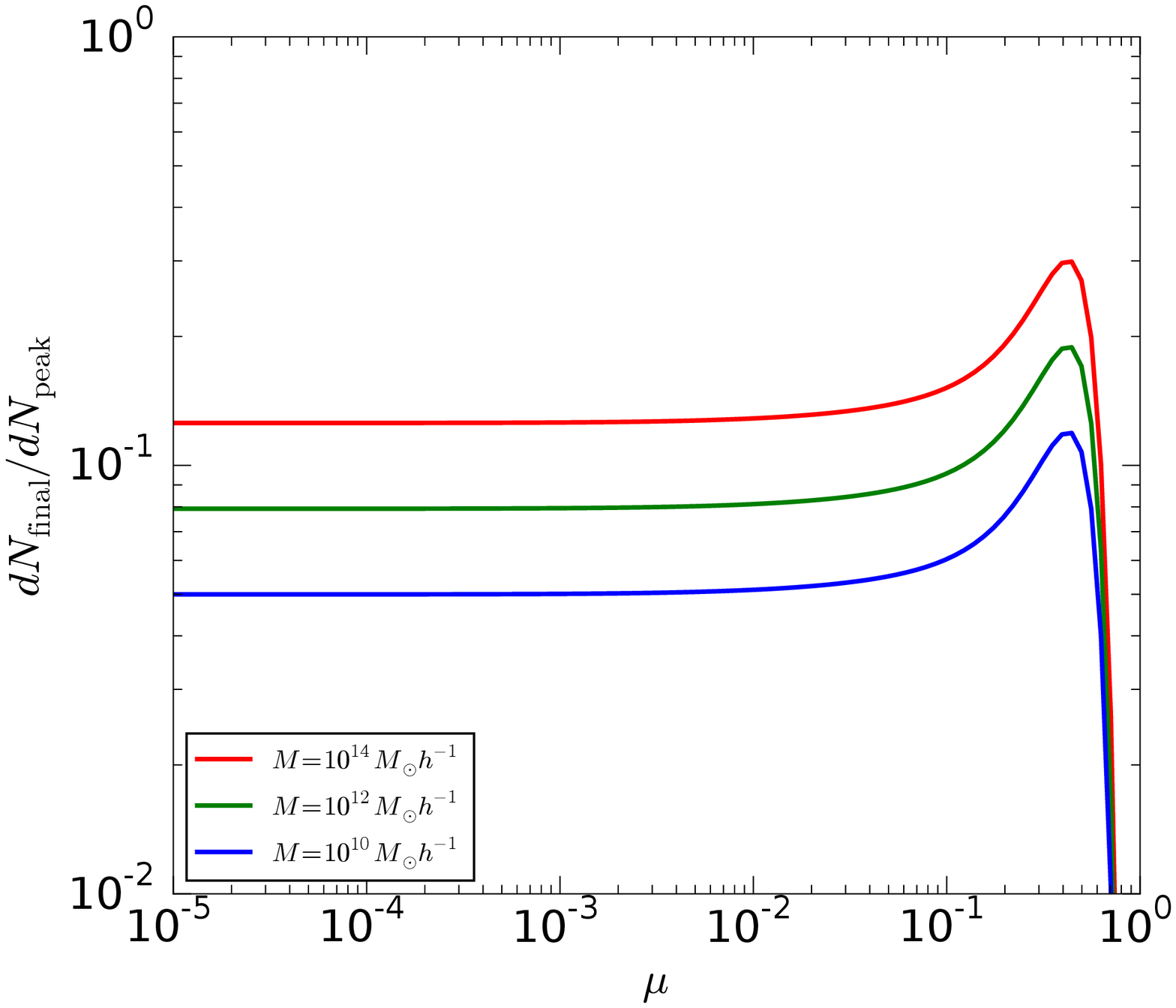}
 \caption{The ratio between the final mass  and peak-mass subhalo mass
   functions, for different host halo masses ($10^{14}$,
   $10^{12}$ and $10^{10}\, \msunh$ from top to
   bottom).}\label{fig:mf_rat} 
\end{figure}

Despite the apparently different parameter values, the peak-mass function depends only weakly on the virial definition. Consistent with previous studies~\citep{vdB05, G08}, the peak-mass function is independent of the host halo mass. On the other hand, the final mass function scales with the host halo mass as $M^{0.1}$ in the host mass range probed by our simulation. This is consistent with the expectation that more massive halos are younger, thus possessing a higher mass fraction in subhalos. 
Overall, the peak mass and final mass functions have similar shapes, while
the presence of a shoulder at $\mu>0.1$ is more prominent in the final
mass function. In Fig.~\ref{fig:mf_rat} we show the ratio between the
two fitted mass functions. For Milky Way sized and cluster sized
halos, the ratio is around $0.1$ at the low mass end, consistent with
the findings of \citet{SubGen}. If tidal stripping of the subhalos is
independent of subhalo mass, then the peak mass and final mass functions are expected to have the same shape. At the high mass end, however, there is a peak in the ratio, indicating that the massive satellites are less stripped than the low mass ones. This is not surprising because when the satellite mass is comparable to the mass of the host halo, the tidal force from the host becomes less important compared with the self-gravity of the satellite.

\subsection{The radial distribution of subhalos}\label{sec:radial}
\begin{figure}
 \myplot{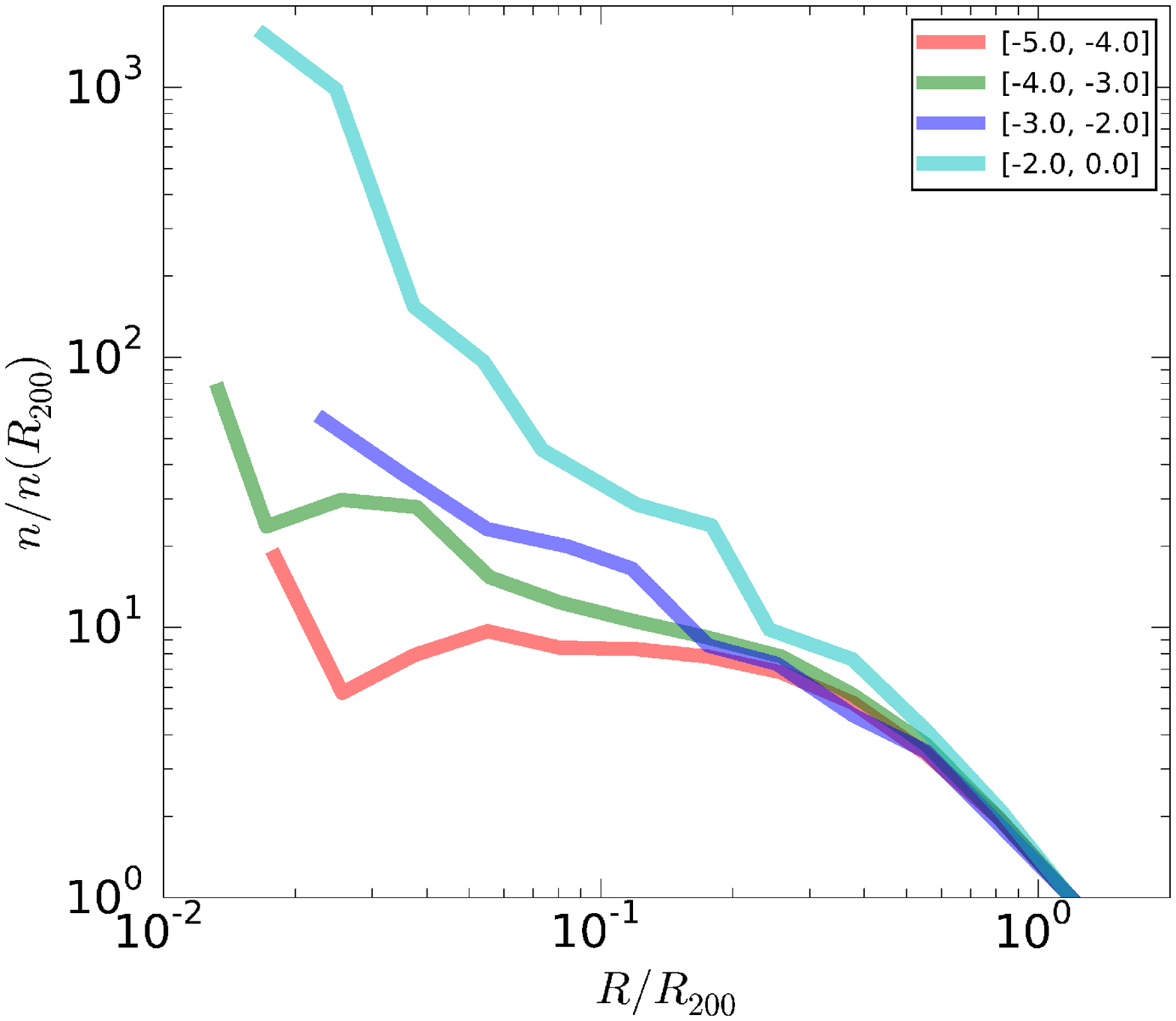}
 \caption{The radial distribution of subhalos in host halos of $10^{13}-10^{14}\, \msunh$. The subhalos are binned in $\log (m/M) $ as labelled. The profiles are normalized by their values at the host virial radius, $R_{200}$. For reference, the gravitational softening of the simulation is about $0.002\, R_{200}$ for these host halos.}\label{fig:radialprof}
\end{figure}

The peak in Fig.~\ref{fig:mf_rat} can be further understood by
reference to the spatial distribution of the subhalos. As shown in
Fig.~\ref{fig:radialprof}, the radial distributions of subhalos of
different relative mass have the same shape near the virial radius,
where the subhalos are barely affected by tidal stripping and are
expected to follow the host halo density profile~\citep{SubGen}. At
smaller radii, however, massive subhalos have a steep profile while
less massive subhalos are depleted at the centre. This pattern is a
consequence of both dynamical friction and tidal stripping. The former
is more important for massive subhalos, making them sink to smaller
radii. At the same time, tidal stripping is less effecient for massive
subhalos, and even more so at the centre of the host when the
satellite largely overlaps with the host, thus failing to eliminate
these objects.

Being free from tidal stripping, the relative abundance of subhalos
near the virial radius is expected to follow the peak-mass function,
i.e., $n(R_{200}, m) \propto f_{\rm peak}(m)$. The subhalo mass
function within the virial radius is simply an integral of subhalo
abundance inside $R_{200}$,   
\begin{align}
 f_{\rm final}(\mu)&=\int_0^{R_{200}} n(r, m) \D^3r\\
 &=n(R_{200}, m) \int_0^{R_{200}}  \frac{n(r, m)}{n(R_{200}, m)} \D^3 r\\
 &\propto f_{\rm peak}(m) \int_0^{R_{200}} \frac{n(r, m)}{n(R_{200}, m)} d^3r.\label{eq:prof_int}
\end{align}
The cuspier radial profile for massive subhalos leads to an increase
in the integral of the relative profile in
Equation~\eqref{eq:prof_int}. As a result, the ratio between the final
mass and peak-mass profiles is also higher for massive subhalos. Note
that the data in \citet{SubGen} only probes the subhalo distribution
at $m/M <10^{-3}$ where this effect is smaller and further suppressed
by the use of a different subhalo finder, as we show explicitly in
Section~\ref{sec:subfind} below. In a follow-up paper, we will extend
the model of \citet{SubGen} to study these distributions in detail. 

\subsection{Comparison with other works}
\subsubsection{Direct comparison with \subfind}~\label{sec:subfind}
\begin{figure*}
\myplottwo{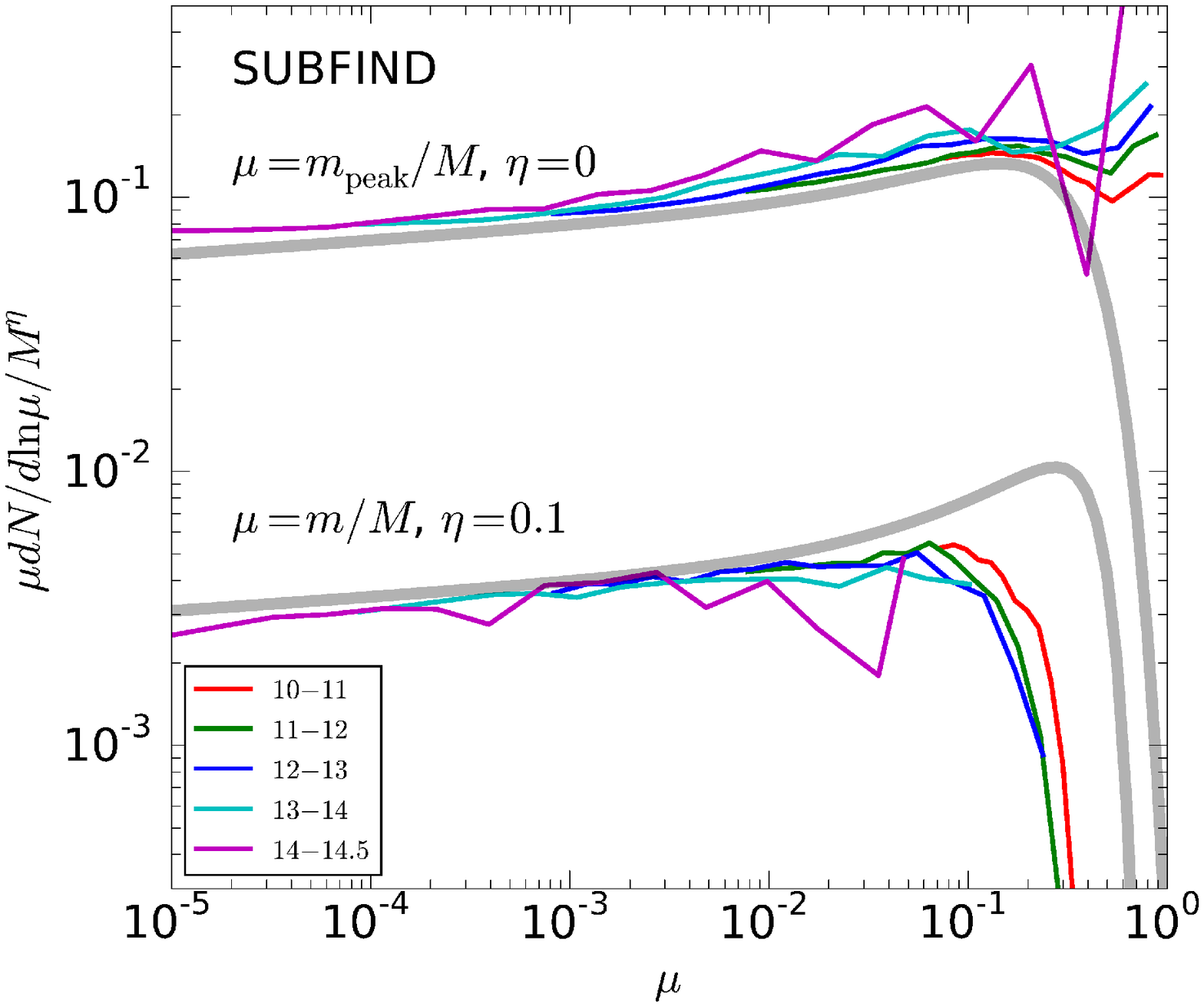}{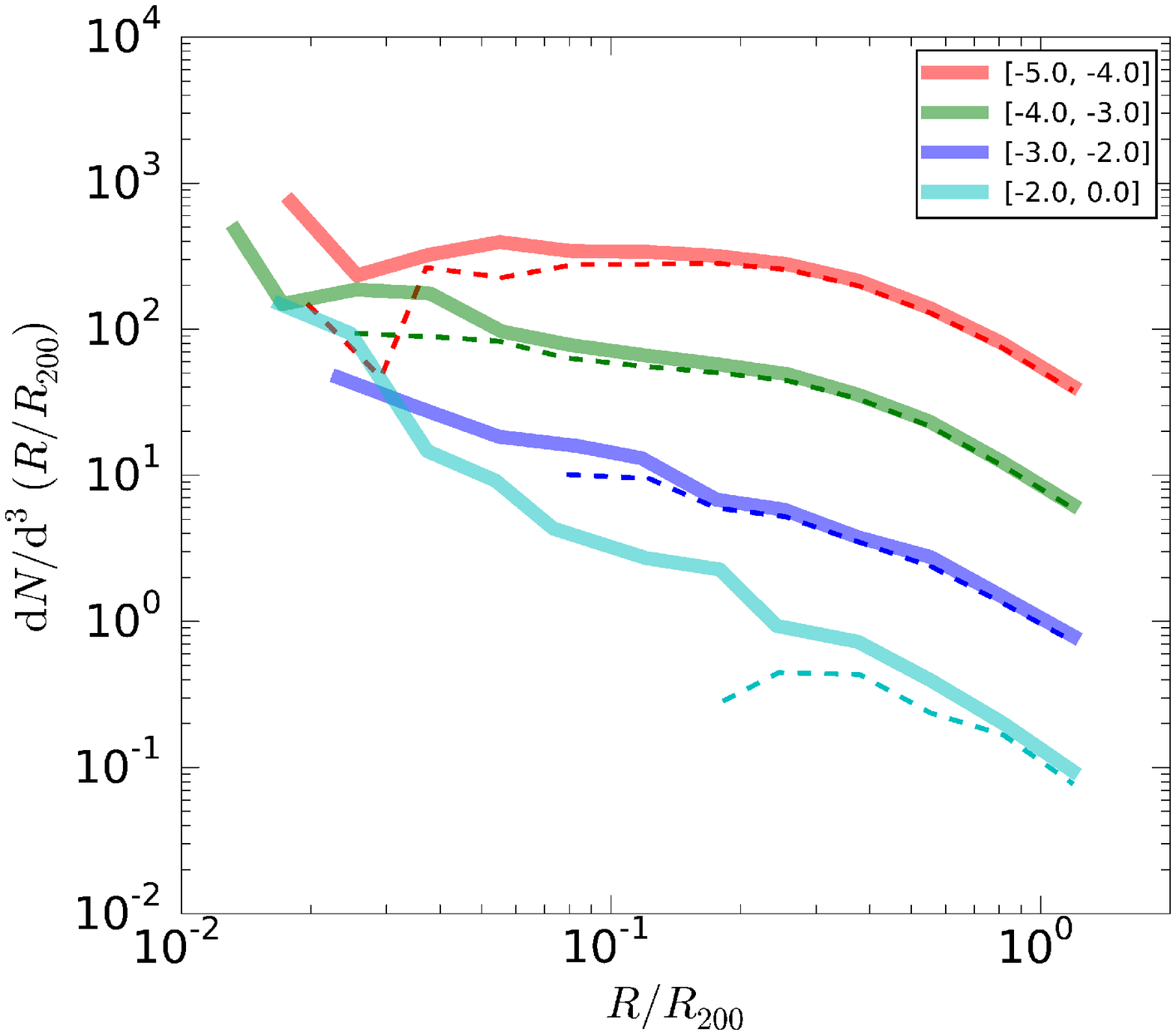}
\caption{Left: The peak mass and final subhalo mass functions for \subfind and \dtree. The thick grey lines are fits to that of \hbtp subhalos (same as in Fig.~\ref{fig:Mill2MF}). The thin lines with different colours are \subtree results in host halos of different masses (labelled by $\log(M/M_\odot h^{-1})$).  Right: the radial profile of \subfind subhalos (dashed lines) compared with that of \hbtp subhalos (solid lines). The host halo mass and the subhalo mass bins are identical to those in Fig.~\ref{fig:radialprof}, except that the profiles are not normalized at $R_{200}$ in order to compare the relative amplitude of the two datasets.}\label{fig:subfind}
\end{figure*}

In Fig.~\ref{fig:subfind} we compare our mass functions and radial
distributions with that found by \subfind~\citep{subfind}. To compute
the peak-mass function, we use the merger tree built by the Durham
merger tree code \dtree~\citep{DHalo}. \dtree has been developed with
efforts to overcome halofinder pitfalls that could cause missing links
or frequent switching of links, and is the default $N$-body merger
tree used by the \textsc{galform} semi-analytic models of galaxy
formation~\citep{galform}. We extend the merger tree by tracking the
most bound particle after termination of a branch, so that the
peak-mass function can be computed in the same way as our code.

As in \hbt~\citep{HBT}, our final mass function is $\sim 10\%$ above
that of the \subfind at the low mass end. At the high mass end,
however, the difference is dramatic, corresponding to a factor of
$2-3$ difference in subhalo mass or an order of magnitude difference
in abundance. This has been pointed out as a problem of \subfind
through comparison with some other halo finders (including \rockstar
and \surv, \citealp{BJ16}) for these massive subhalos.  

In the right panel of Fig.~\ref{fig:subfind}, we explore this
difference in the spatial distribution of subhalos. In the outer halo,
our subhalos are slightly more abundant, which can be understood as
our subhalos being slightly more massive. The overall shape of the
distributions are still quite consistent with each other. In the inner
halo, however, \subfind shows a deficiency of subhalos compared with
our result, which is most significant for more massive subhalos. This
can be understood as a reflection of the `blending problem' exhibited
by configuration space subhalo finders: when a subhalo overlaps with
the host halo, it is difficult to separate it from the host using
only density information. It is easy to understand that this issue
is more severe for larger subhalos. 

The deficiency of massive subhalos near the centre of the host halo in catalogues constructed using \subfind explains, at least in part, the disagreement noted by~\cite{Schwinn} and~\cite{HSTFF} between the distrubtion of subhalos identified using \subfind in \LCDM simulations and the distribution of subhalos inferred from lensing studies in galaxy clusters. \cite{Mao_17} have argued that the mismatch between the simulation result of Millennium-XXL~\citep{Angulo_12} and the lensing results of \citet{Jauzac16, Schwinn} is also affected by the use of different masses in the comparison: \subfind masses in the simulation but projected aperture masses from lensing that also include mass contributions from the host. Even with the use of deblended subhalo mass estimates in the strong lensing analysis of \citet{HSTFF}, however, the observed spatial distribution of massive subhalos is still found to be more centrally concentrated than that from the \subfind catalogues of the Illustris simulations~\citep{Illustris}, which can be attributed to this radial-dependent blending issue in the \subfind catalogues. 

In contrast to the final mass function, the peak-mass function from
\subtree is systematically above ours. Together with the lower
final mass function, this means more branches are produced in the
merger tree of \subtree. As we will show explicitly in
Section~\ref{sec:persistency}, this difference can be attributed to
broken branches associated with missing links and the switching of
subhalo masses in \subtree. 

\subsubsection{Comparison with fitting functions and the low mass end slope} 
\begin{figure*}
 \myplottwo{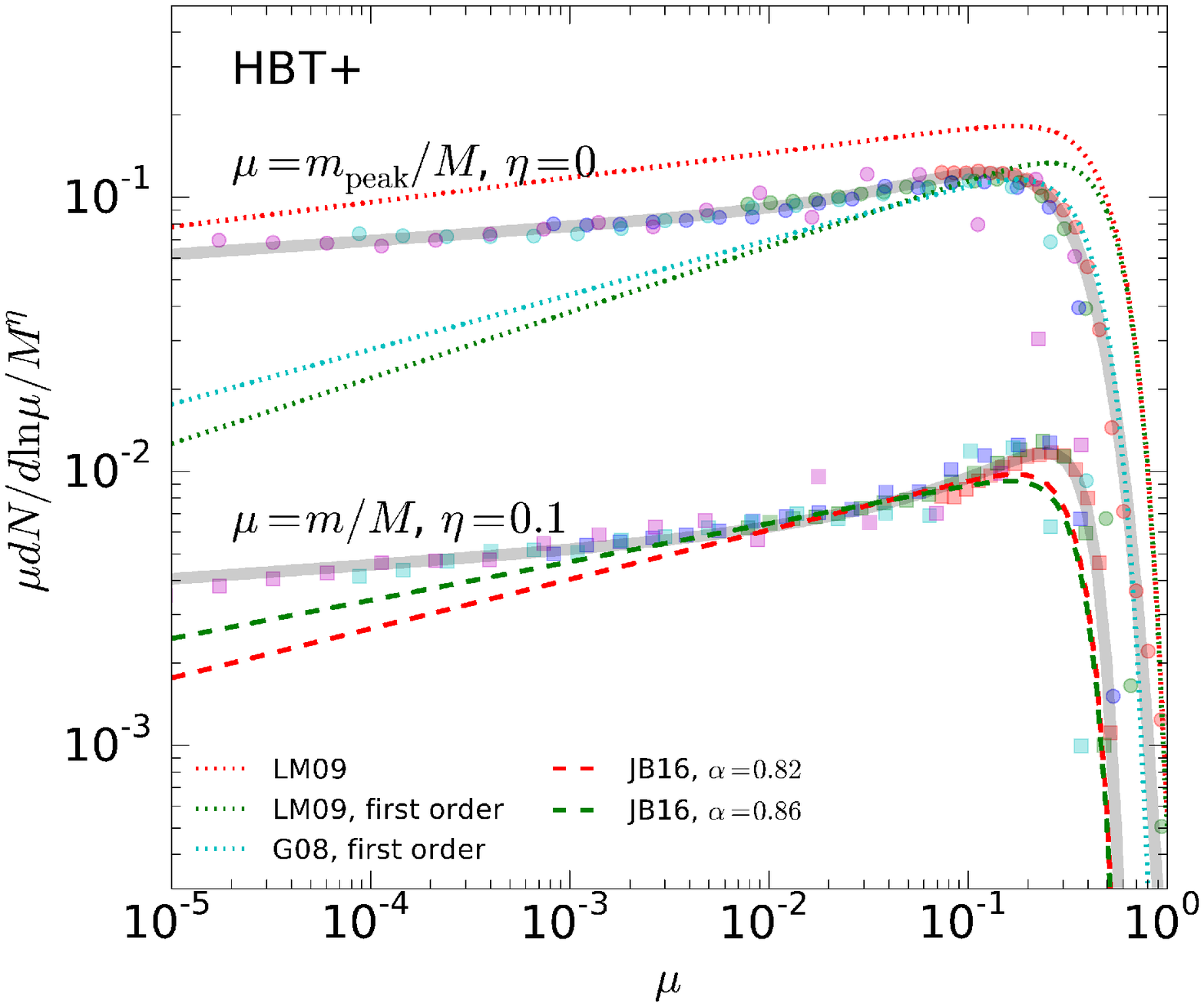}{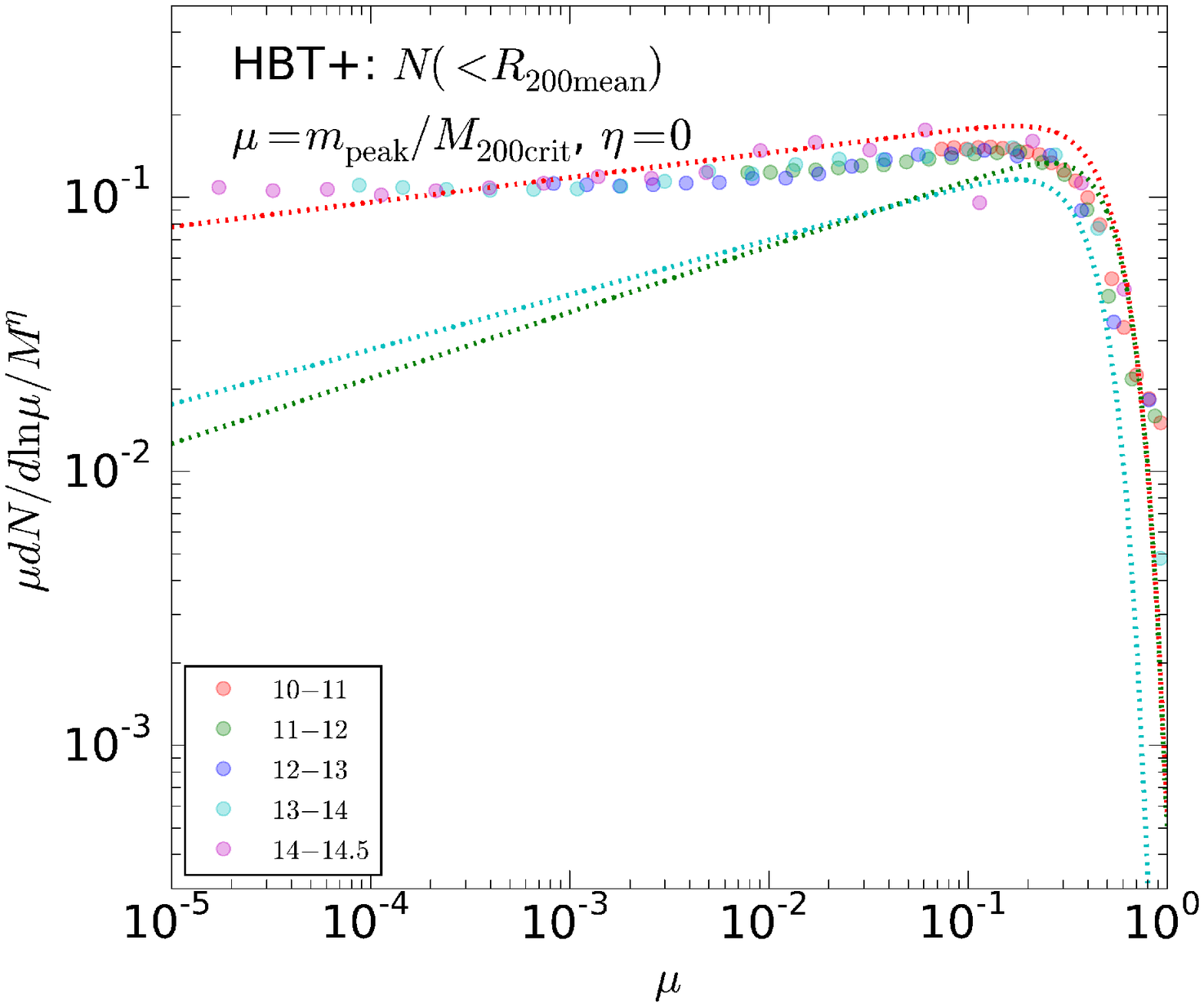}
 \caption{The subhalo mass functions compared with fitting functions
   in previous works. As in Fig.~\ref{fig:Mill2MF}, the data
   points are results from our code, with different colors
   representing different host halo mass bins. The squares and circles
   show the peak mass and final subhalo mass functions respectively.  The
   thick lines are fits to the data points as listed in
   Table~\ref{table:par}. The peak-mass function is compared with the
   fitting functions of \citet{LM09} for all the subhalos and
   first-order subhalos, as well as with the fit of \citet{G08} for
   first-order subhalos. The final mass function is compared with the
   fitting function of \citet{JB16} and \citet{BJ16} which are
   calibrated using the  ROCKSTAR halo finder. In the left panel, we adopt the
   ``Virial'' definition for both  host mass and
   radius. In the right panel, the host halo mass is defined according
   to  the ``200Crit'' definition while the host halo radius is defined
   according the ``200Mean'' definition.}\label{fig:compJB}
\end{figure*}

In Fig.~\ref{fig:compJB} we compare our results against a few fitting
functions in the literature. In the left panel, we have switched to
the same definition of virial  quantities (``Virial'' as in Table~\ref{table:par})
as in \citet{G08, JB16, BJ16} when computing the mass functions. The
model of \citet{JB16} is a semi-analytical model that evolves
progenitor halos generated from extended Press-Schechter merger trees
according to an empirical average mass stripping rate. Their model is
calibrated against \rockstar subhalos from the Bolshoi~\citep{Bolshoi}
and MultiDark~\citep{Multidark} simulations, and the resulting subhalo
mass function is fitted using a Schechter function. The normalization
of the mass function is predicted from the dynamical age of the host
halo, which we find can be equivalently fitted with a power-law in the
mass range $10^{10}-10^{15}\msunh$ for the concordance cosmology,
consistent with our scaling in Table~\ref{table:par}. For the purpose of comparison with our results, the final
fitting function of the \citet{JB16} model can then be summarized as
\begin{equation}
 \frac{\D N}{\D \ln \mu}= a \left(\frac{M}{10^{10} \msunh}\right)^{\eta}  \mu^{-\alpha} \exp \left( - b \mu^{\beta} \right),\label{eq:mf_jiang}
\end{equation} with $a=0.014$, $\eta=0.1$, $\alpha=0.82$, $b=50$, $\beta=4$ according to \citet{JB16}, and slightly different parameters $a=0.012$, $\alpha=0.86$ in \citet{BJ16}. Our results are quite consistent with their fitting function with $\alpha=0.86$ in the mass range $10^{-3}<\mu<10^{-1}$. At the high mass end, our data shows a higher shoulder around $\mu=0.3$, indicating that the mass stripping rate of the most massive subhalos differs from the average mass stripping rate of low mass ones in the framework of the \citet{JB16} model. 
At the very low mass end, our data clearly support a slope higher than $\alpha=0.86$. While \citet{BJ16} argued that their data is in contradiction with a low mass end slope of $\alpha=0.95$, our results suggest that their conclusion is caused by the limited mass range in their data ($\mu>10^{-3}$). The local slope is an increasing function of subhalo mass due to the decrease in tidal stripping efficiency at the high mass end, and the asymptotic $\alpha$ is still consistent with $0.95$ down to $\mu\sim 10^{-5}$.

For the peak-mass function, \citet{G08} has measured the first-level unevovled mass function, that is, the mass distribution of progenitors that fall directly into the host halo (instead of being accreted as a satellite of another infalling halo). At the high mass end, the peak-mass function is dominated by these first-level progenitors, and our measurement agrees very well with that of \citet{G08}. At the low mass end, higher level contributions become more important, and the \citet{G08} fit falls below our measurement. \citet{LM09} measured both the first-level and all level unevolved mass function, which have been used in \citet{Jiang14,JB16} as benchmarks to calibrate Monte-Carlo merger trees from extended Press-Schechter theories. Their results lie mostly above our measurements. However, it should be noted that \citet{LM09} adopted a somewhat peculiar combination of the mass and spatial extension of a host halo. Their merger trees are based on Friends-of-Friends (FoF) halos, while the host halo mass is defined as $M_{\rm 200Crit}$ corresponding to a spherical overdensity that is 200 times the critical density of the universe. To make a better comparison, in the right panel of Fig.~\ref{fig:compJB} we compute the peak mass function inside $R_{\rm 200Mean}$, which is expected to be closest to the size of a Friends-of-Friends halo, while adopting $M_{\rm 200Crit}$ as the host halo mass. Such a combination leads to a mass function that is closer to the \citet{LM09} results, but still lower at the high mass end. This can be further attributed to the fact that they rely on Friends-of-Friends halos to build their merger trees. In this case, halos that are temporarily linked together and subhalos that have been ejected from the host can both contribute to the progenitor mass function. The situation can become even worse if these objects fall back into the host and are counted multiple times~\citep{Benson16}. In contrast, our approach of selecting branches located inside the final virial radius produces a progenitor population that can be unambiguiously compared against the final mass function inside the virial radius. Due to these complications, the \citet{LM09} results should be quoted with caution in future analytical studies.


\subsection{The persistence of tracks}\label{sec:persistency}
\begin{figure}
 \myplot{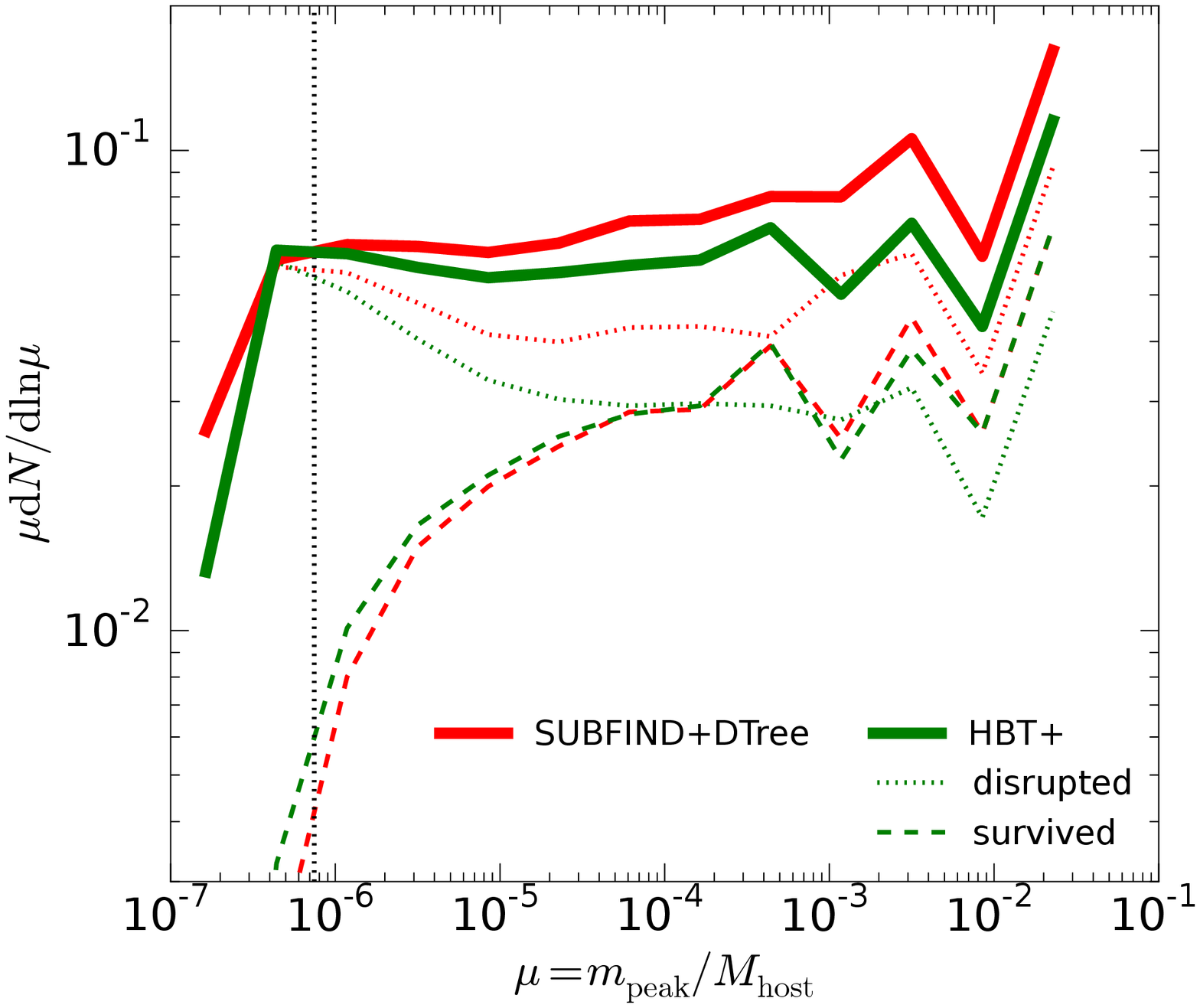}
 \caption{The peak-mass function (solid line) of tracks in Aquarius halo A2, decomposed into disrupted and surviving populations (dotted and dashed lines). The red and green colours show the results of \subtree and \hbtp respectively.}\label{fig:A2MF}
\end{figure}

To further investigate the difference in the peak-mass function
between \subtree and \hbtp, we carry out a detailed study
focusing on a single high resolution halo, AqA2 from the Aquarius
simulation set. Fig.~\ref{fig:A2MF} shows the peak mass computed from
our code and that from \subtree. Consistent with
Fig.~\ref{fig:Mill2MF}, the peak-mass function from \subtree is
higher than ours. According to whether the descendant subhalo at $z=0$
is still resolved, we decompose the peak-mass function into a
surviving and a disrupted component. The peak-mass functions of
surviving subhalos 
agree well with each other, meaning that both codes have identified
the same population of final subhalos. On the other hand, the
disrupted peak-mass functions can differ up to a factor of $2$, with
\subtree having more disrupted branches.  

\begin{figure}
 \myplot{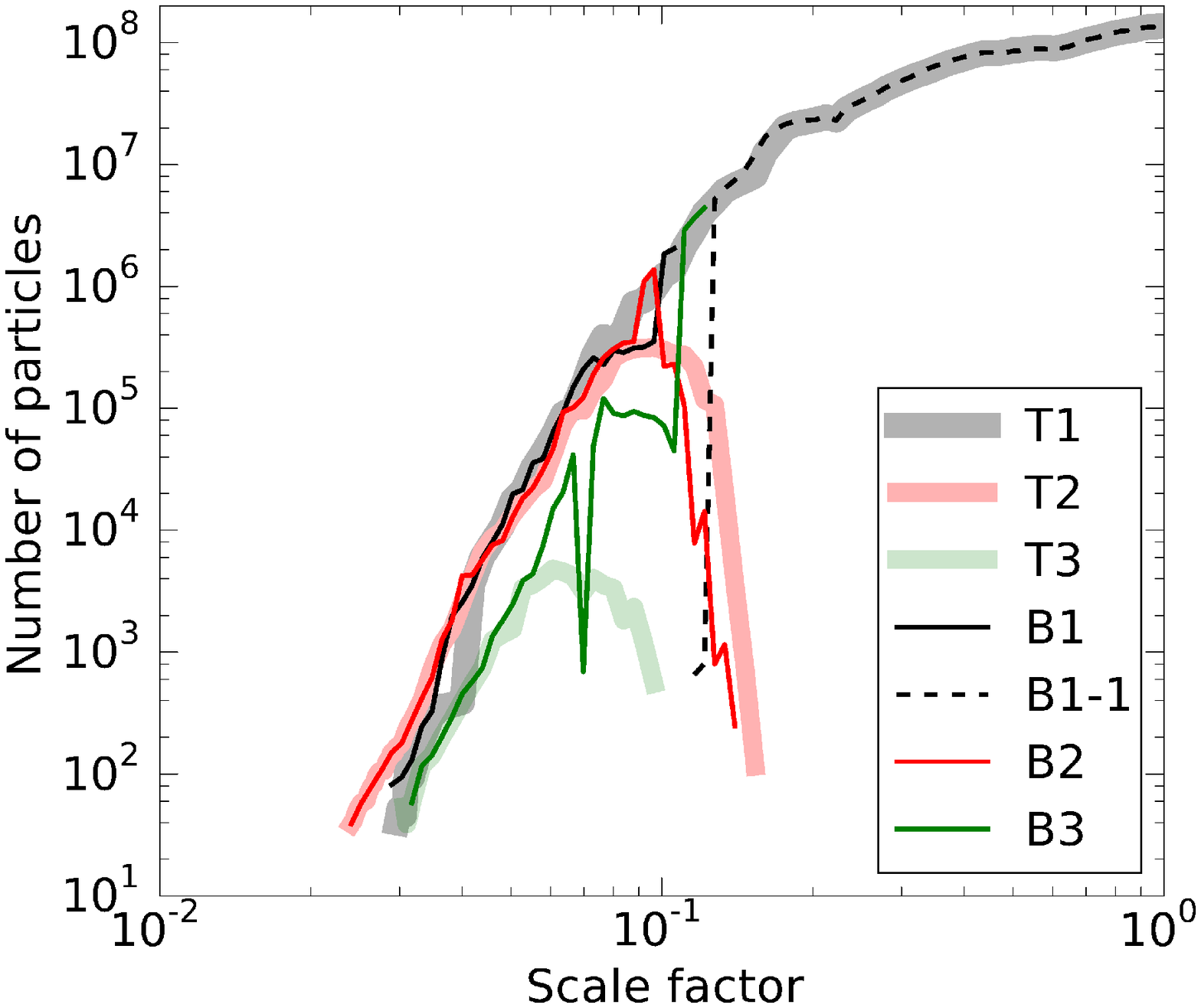}
 \caption{The mass evolution of the central subhalo of AqA2. The thin lines show the mass evolution history of subhalos that are identified as the central subhalo of AqA2 at different times according to \subtree. The thick lines show the corresponding branches as identified by \hbtp. }\label{fig:A2MAH}
\end{figure}

\begin{figure*}
 \myplottwo{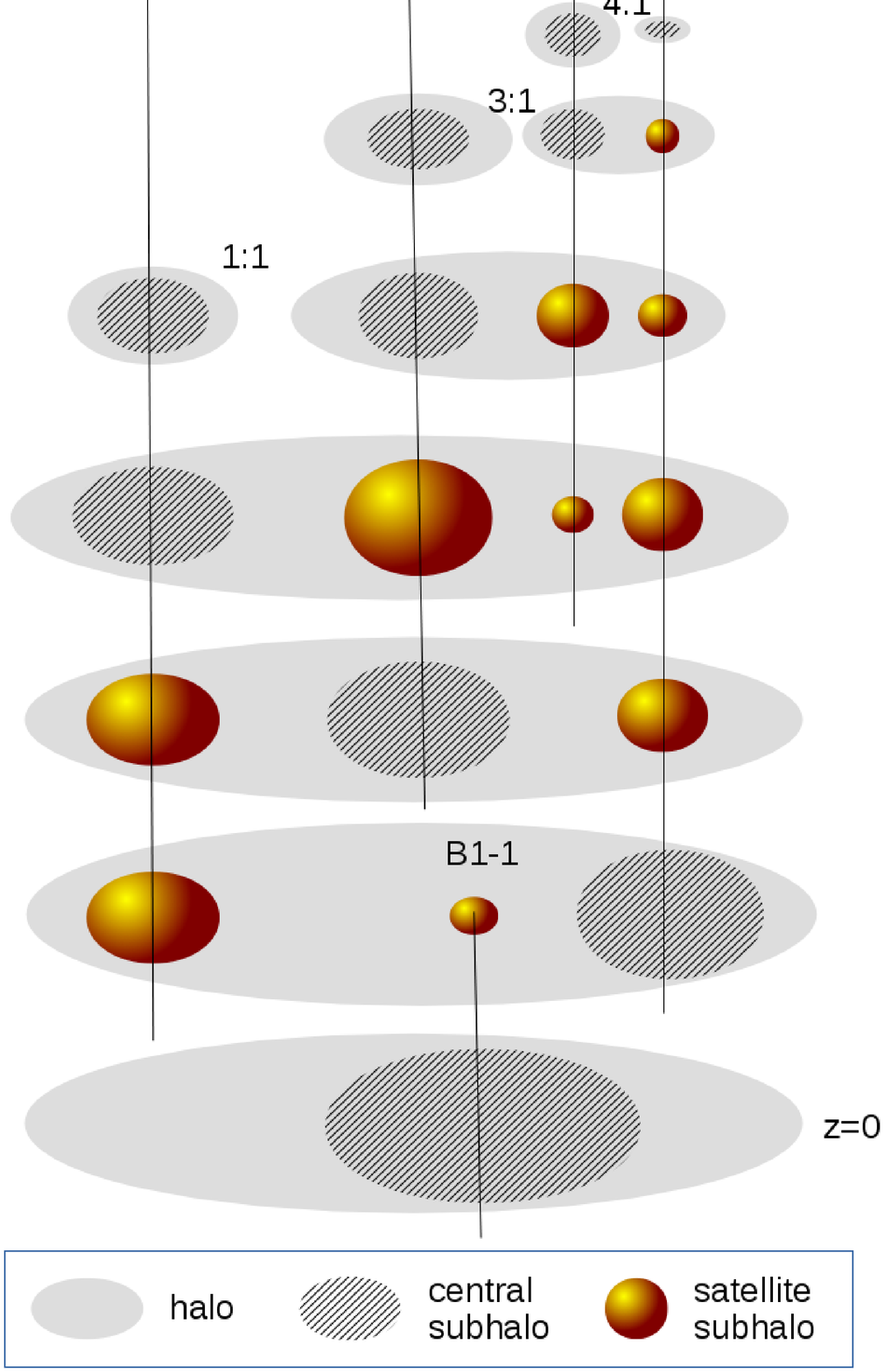}{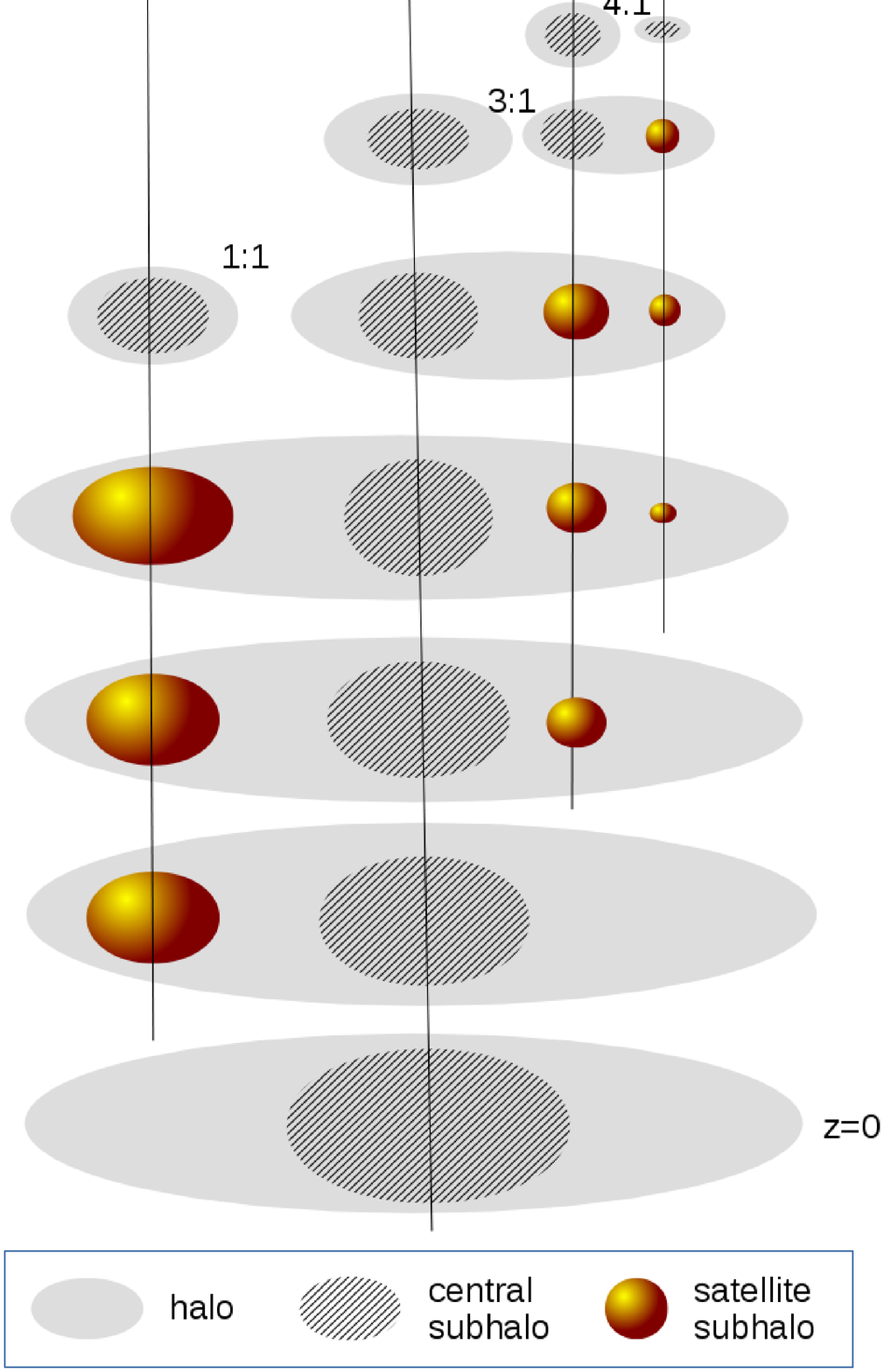}
 \caption{The major merger history of halo AqA2 as resolved by \subtree (left) and \hbtp (right). The ratios next to halos list the mass ratio of the progenitor halos just before the merger. Each track (vertical lines) terminates after the disruption of the subhalo. For illustration purpose, only the snapshots of interests are plotted.}\label{fig:A2Tree}
\end{figure*}

In the $(r, m_{\rm peak})$ plane, we have identified some of these
extra branches that exists in \subtree but not in \hbtp.
Fig.~\ref{fig:A2MAH} shows the mass evolution history of the three
most massive branches (B1, B2, B3) selected this way. Correspondingly,
we have also identified branches in \hbtp that best match the selected
branches in orbital and mass evolution. Interestingly, these branches
are related to two major merger events that happened to the host halo.
In the \subtree case, the B1, B2, and B3 branches all
temporarily become the most massive branch in the host at some stage,
and then get disrupted almost immediately after that. The final
central subhalo emerges abruptly from inside the host halo, as shown
by the B1-1 branch. In contrast, the corresponding T2 and T3 branches
in \hbtp remain as less massive branches than T1 after merger until
they are fully disrupted. The T1 branch remains as the most massive
branch in the host halo untill the final time. In
Fig.~\ref{fig:A2Tree}, we visuallize this evolution as a track table,
where an additional branch, B4 (T4), is included to show the major
merger with B3 (T3) that leads to the strange mass growth in B3. In
the \hbtp case, the central and satellite subhalos are tracked
consistently and persistently, while the \subtree tree suffers
a few switches in the mass and in the central-satellite determination,
as well as a broken link that fragmented B1-1 from B1. The switching 
problem leads to an overestimate of the peak mass, while the broken
links create extra progenitor branches. Both of these lead to an
overestimate of the peak-mass function.

\section{Summary and conclusion}~\label{sec:summary}

We have presented an improved version of the original \hbt algorithm
of~\citet{HBT} that tracks halos through time to find subhalos and build merger trees. 
A series of improvements are implemented,
including: \begin{itemize}
 \item Treatment of subhalos as Lagrangian objects and organization of
 the merger tree as a table of tracks. This allows intuitive and
 flexible storage and retrieval of subhalos and trees. 
 \item Significant improvement in speed. This is made possible  by a
   physically motivated yet simple algorithm for the identification of
   the main progenitor halo and a refined unbinding algorithm with a
   complexity of $O\left(N\right)$ compared to the $O\left(N\log
     N\right)$ complexity of a plain tree code.
 \item Detection and merging of trapped subhalos. These are  massive
   satellites that  sink to the centre of their host halo and remain 
   there without being disrupted, leading to pairs of subhalos that overlap
   in their orbit while remaining individually self-bound. We have
   developed a prescription to detect such pairs and merge the trapped satellite.
 \item Support for distributed computation through \mpi.
 \item Support for hydrodynamical simulations.
\end{itemize}
The code has been rewritten in \textsc{C++} with user friendly
configuration options, as well as \textsc{HDF5} output format that
allows direct postprocessing with other software. The 
source code is publicly available at \url{https://github.com/Kambrian/HBTplus} and
\url{http://icc.dur.ac.uk/data/}.

As an illustration we applied the new code to a study of the
distribution of subhalos and tested the persistence of merger trees in
the Millennium-II simulation and in one of the Aquarius project
simulations. In contrast to previous studies that fit the mass
functions with a single Schechter function, we find that both the
final and peak-mass subhalo mass functions are well fitted by a double
Schechter function (Eq.~\eqref{eq:mf}) with similar shapes. These mass
functions harden towards the high mass end before falling off
exponentially. The hardening is most significant in the final mass
function, reflecting an inefficiency of tidal stripping of massive
subhalos. This also reflects our finding that the radial distribution
of massive subhalos is more concentrated than the universal radial
profile of low mass ones, due to stronger dynamical friction and
weaker tidal stripping. The detection of this hardening requires the
ability to identify subhalos in the inner regions of the host halo,
which \hbtp does, but which subhalo finders that work in configuration
space alone have difficulty identifying. 
Recent lensing observations of galaxy clusters have resulted in
reports of discrepancies in the observed subhalo distribution 
compared to that of \LCDM predictions, including the excess of massive subhalos reported
by \citet{Jauzac16} and \citet{Schwinn} and the more concentrated subhalo radial 
distribution reported by \citet{HSTFF}. These discrepancies can be explained, 
at least in part, by the blending problem present in the Millennium-XXL and
Illustris \subfind catalogues used in their comparisons.

The hardening of the subhalo mass function at the high mass end means
that single power-law fits to the mass function are inadequate. From
our \hbtp subhalo catalogues constructed from the Millennium-II
simulation, we find that, when the entire mass function is fitted with
a double Schechter function, the low-mass end slope (down to $m/M_{\rm
  host}=10^{-5}$) is consistent with a power-law exponent,
$\alpha=0.95$. 

We have demonstrated that the peak-mass function, or the ratio between
the peak mass and final mass functions, are good statistics to test
the quality of merger trees. The existence of broken or false links in
the trees introduces extra branches and inflates the peak-mass
function, which can be overestimated by as much as a factor of 2 in an
complex merger tree built from \subfind subhalos, as a result of the
`blending' and `mass or centre-switching' problems that are present in
the latter. This issue is important for studies that focus on the
remnants of the most massive progenitors in a halo, such as studies of
streams in the Milky Way halo. It also has important implications for abundance matching models that match galaxies to simulated merger trees using peak masses: the peak mass function inflated by broken and false links could lead to false matches of galaxies to broken branches, subsequently biasing the inferred properties of massive satellite galaxies. In contrast, our algorithm is robust against these problems by design. It is able to track the tree branches persistently, and recovers a higher final mass function, as well as a lower and universal peak-mass function. 


\section*{Acknowledgments}
JXH benefited from fruitful discussions with Yipeng Jing, Peter Berhoozi, Tom Theuns, Matthieu Schaller, Surhud More, Mathilde Jauzac and Marius Cautun, and is grateful to Tianchi Zhang, Xianguang Meng and Zhaozhou Li for providing user feedback during the development of the code. Kavli IPMU was established by World Premier International Research Center Initiative (WPI), MEXT, Japan. 
This work was supported by the Euopean Research Council [GA 267291] COSMIWAY, Science and Technology Facilities Council
Durham Consolidated Grant, and JSPS Grant-in-Aid for Scientific Research JP17K14271. 
This work used the DiRAC Data Centric system at Durham University,
operated by the Institute for Computational Cosmology on behalf of the
STFC DiRAC HPC Facility (www.dirac.ac.uk). This equipment was funded
by BIS National E-infrastructure capital grant ST/K00042X/1, STFC
capital grant ST/H008519/1, and STFC DiRAC Operations grant
ST/K003267/1 and Durham University. DiRAC is part of the National
E-Infrastructure. This work was supported by the Science and
Technology Facilities Council [ST/F001166/1]. 

\bibliographystyle{\mybibstyle}
\setlength{\bibhang}{2.0em}
\setlength\labelwidth{0.0em}
\bibliography{ref}

\appendix
\section{Sampling noise in unbinding}\label{sec:unbind_noise}
\begin{figure}
 \myplot{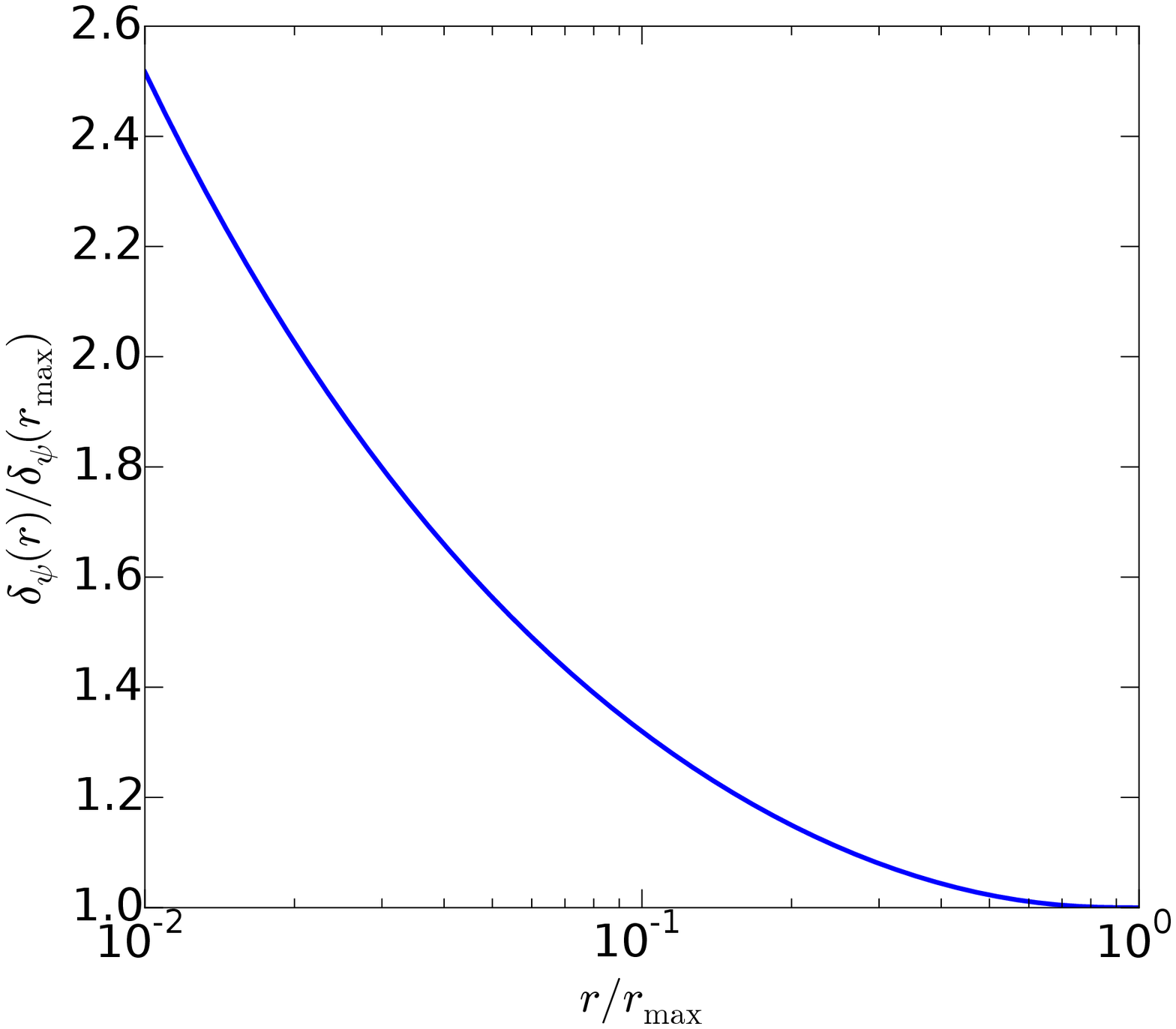} 
 \caption{The radial dependence of the Poisson noise in the estimated potential for an isothermal sphere sampled with particles. The noise increases as the radius decreases.}\label{fig:delta_psi_iso}
\end{figure}

Consider a singular isothermal sphere sampled with $N$ particles out to a truncation radius $r_{\rm max}$. Expressed in the dimensionless radius $\tilde{r}=r/r_{\rm max}$, the cumulative number density profile of the halo particles is $N(<\tilde{r})=N \tilde{r}$. Assuming Poisson fluctuations in the particle counts at each radius, the uncertainty in the estimated potential can be obtained as
\begin{align}
 \delta_{\psi}(r)&\equiv\frac{\sigma_\psi(r)}{|\psi(r)|}\\
 &=\frac{\sqrt{\frac{N(r)}{r^2}+\int_{r}^{r_{\rm max}} \frac{\D N(R)}{R^2}}}{\frac{N(r)}{r}+\int_{r}^{r_{\rm max}} \frac{\D N(R)}{R} }\\
 &=\frac{1}{\sqrt{N}} \frac{\sqrt{2/\tilde{r}-1}}{1-\ln\tilde{r}}.
\end{align} At $r=r_{\rm max}$, the uncertainty is the smallest with $\delta_{\psi}(r_{\rm max})=1/\sqrt{N}$. The radial dependence of this Poisson noise is shown in Fig.~\ref{fig:delta_psi_iso}.

\begin{figure}
 \myplot{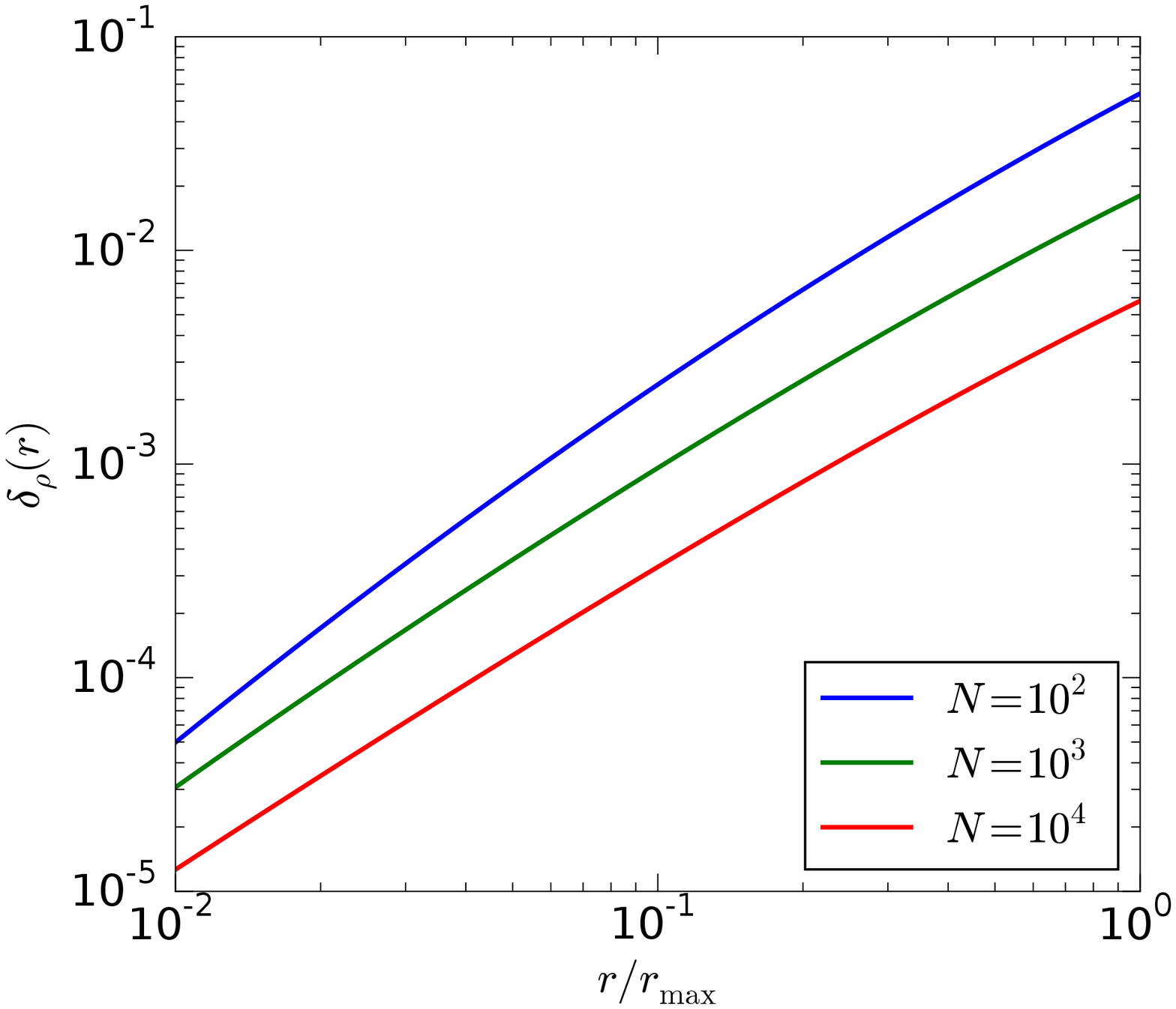}
 \caption{The relative uncertainty in the self-bound density profile due to Poisson noise in the potential, for an isothermal sphere sampled with $N$ particles.}\label{fig:delta_rho_iso}
\end{figure}

The uncertainty of the bound density profile due to Poisson noise can be estimated as
\begin{align}
 \delta_{\rho}(r)&\equiv\frac{\sigma_\rho(r)}{\rho(r)}\\
 &=\frac{\int_{\sqrt{2|\psi|}}^{\sqrt{2(|\psi|+\sigma_\psi)}} \exp\left(-\frac{v^2/2+\psi}{\sigma^2}\right) d^3v}{\int_{0}^{\sqrt{-2\psi}} \exp\left(-\frac{v^2/2+\psi}{\sigma^2}\right) d^3v}\\
 &=\frac{f(b)}{f(a)}-1,
\end{align} where $f(x)=\sqrt{\pi} \erf(x)-2e^{-x^2}x$, $a=\sqrt{|\psi|/\sigma^2}=\sqrt{2(1-\ln\tilde{r})}$, $b=\sqrt{(|\psi|+\sigma_\psi)/\sigma^2}=\sqrt{2(1-\ln \tilde{r}) (1+\delta_\psi(r))}$, and $\sigma^2$ is the one-dimension velocity dispersion of the halo. This is plotted in Fig.~\ref{fig:delta_rho_iso} for a few different sample sizes, $N$. For $N=10^3$, the density profile can be recovered to percent level accuracy or better over the entire radial range.

Even though the potential estimate is less accurate at smaller radii, the unbinding of the particles is almost unaffected by Poisson noise at these radii. This is because the potential is much deeper at small $r$. Given the constant velocity dispersion in the isothermal sphere, most of the particles at a small $r$ are tightly bound, making the unbinding insensitive to the accuracy in potential. This is also true for NFW haloes, in which the velocity dispersion is known to eventually decrease towards the halo centre.

\section{Trapped subhalos}~\label{sec:sink}
\begin{figure}
 \myplot{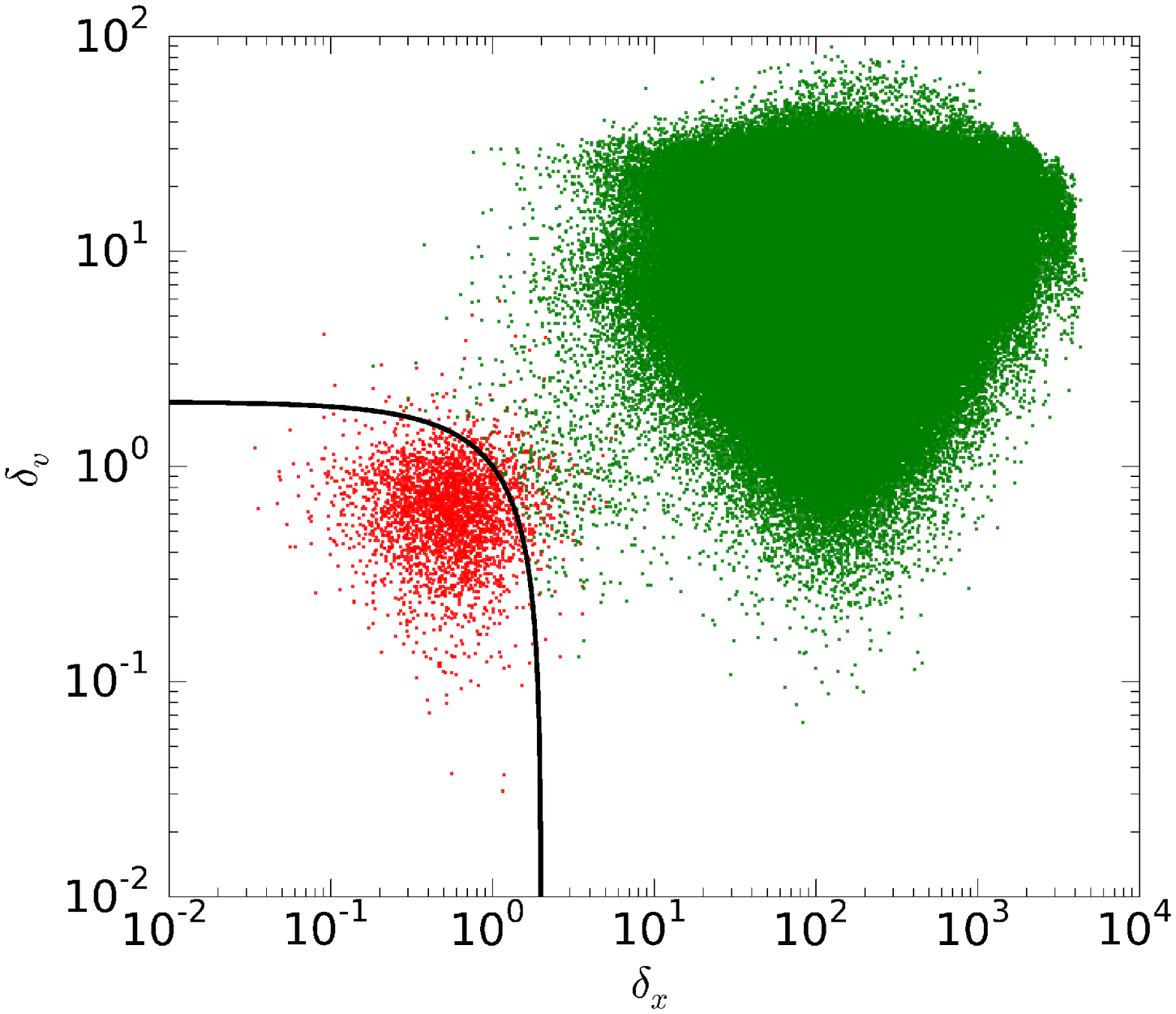}
 \caption{The distribution of satellites in the $(\delta_x,\delta_v)$ plane at $z=0$ in Millennium-II. We select satellites that are still resolved and whose host subhalo has more than $10^4$ particles. $\delta_x$ and $\delta_v$ are the position and velocity offset of the satellite from its host subhalo, normalized by the position and velocity dispersions at the host centre respectively. The red dots are satellites tagged as trapped while the green dots are the remaining ones. The black line marks $\delta_x+\delta_v=2$, the critical curve used to identify trapped subhalos in the merger history.}\label{fig:DeltaDistr}
\end{figure}

Massive satellites are likely to sink to the centre of their host halo without getting disrupted. During this process, the orbital energy of the central-satellite pair is converted to the internal energy of each subhalo, and the satellite is trapped in the centre thereafter. To see that these satellites are indeed a distinct population, in Fig.~\ref{fig:DeltaDistr} we show the distribution of satellites according to their position and velocity offset from their host subhalos. Here $\delta_x=\Delta_x/\sigma_x$, where $\Delta_x$ is the separation of the satellite from its host subhalo, and $\sigma_x$ is the position dispersion of the $20$ most-bound particles in the host subhalo. Similarly, $\delta_v=\Delta_v/\sigma_v$ is the normalized velocity offset. It is obvious that the satellites show a bimodal distribution in this plane, with the trapped subhalos clustered around $(\delta_x, \delta_v)=(1,1)$, consistent with them being draw from the most bound particles in the host subhalo. Note that in this test we have not merged the trapped subhalos in order to make this plot, but only tag them as trapped as soon as they reach $\delta=\delta_x+\delta_v<2$ in their orbital evolution.

\begin{figure}
\myplot{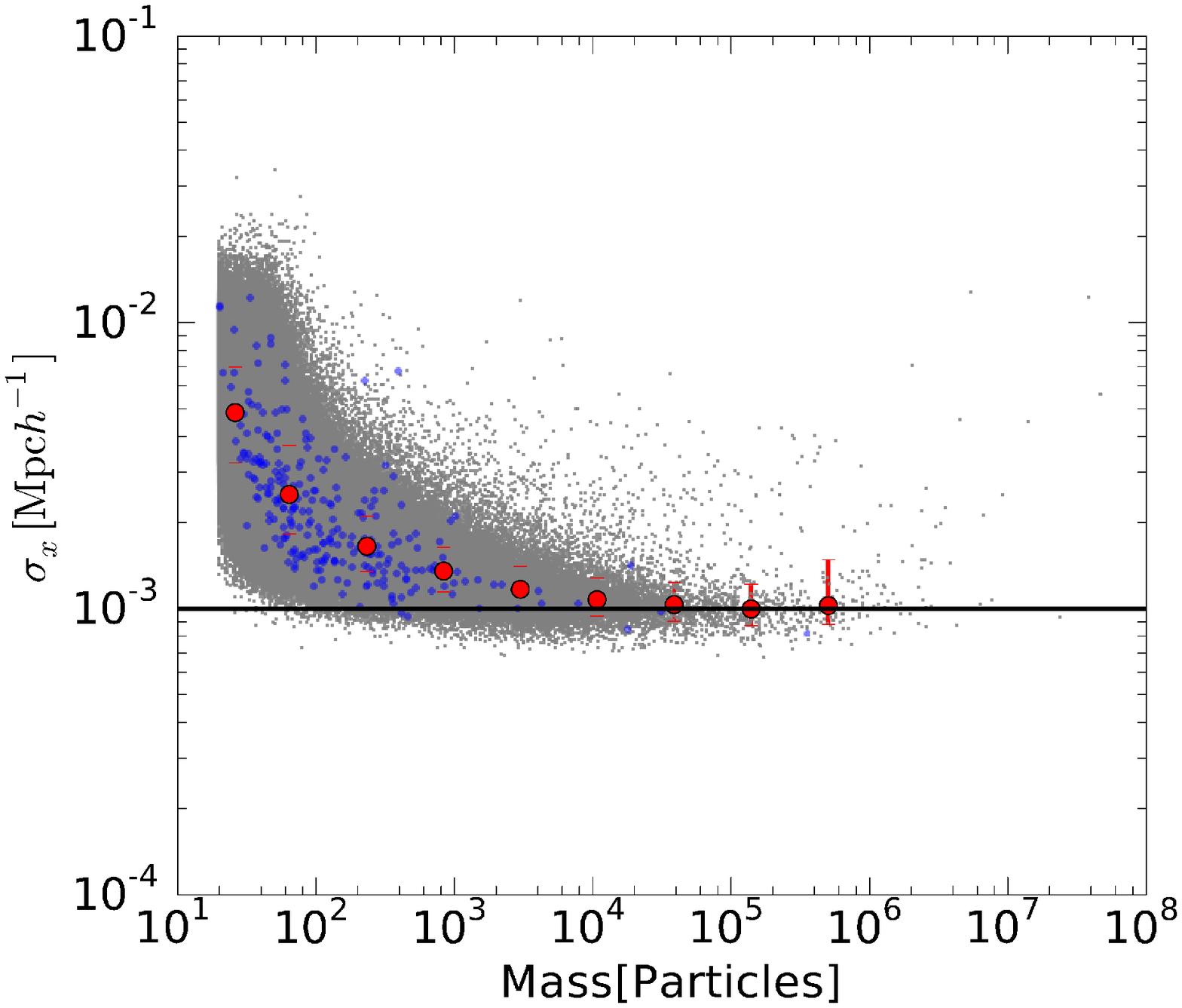}\\
\myplot{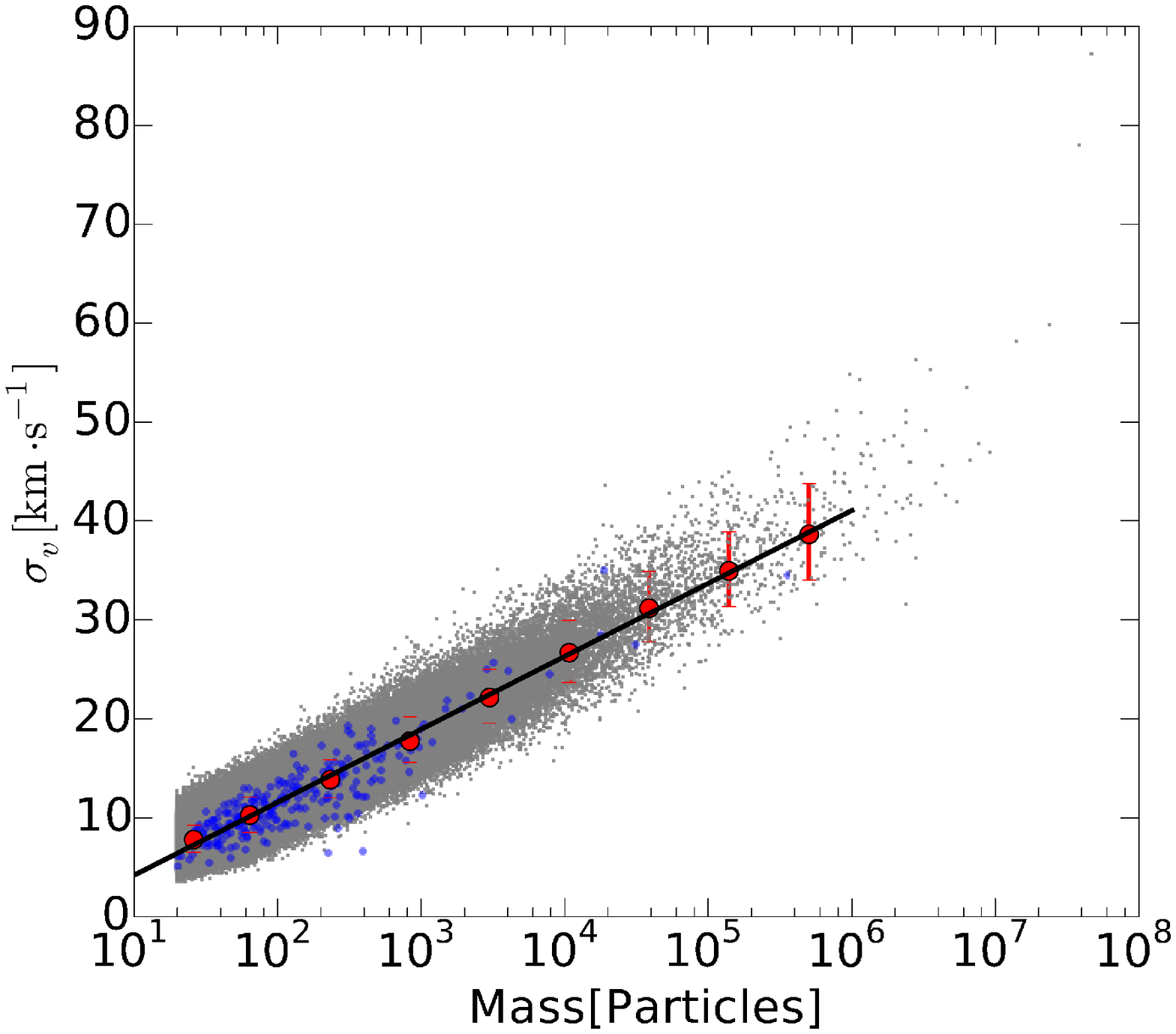}\\
\caption{The position and velocity resolution in subhalos of different masses in Millennium-II. The grey dots are the results for central subhalos while the blue dots are those for satellites. In this figure we use the peak mass of a satellite as its mass. For clarity, only $1/10$ of the central subhalos and $1/10000$ of the satellite subhalos are plotted. Filled circles with errorbars show the median and $\pm 1\sigma$ percentiles of the distributions. In the top panel, the horizontal solid line marks the force softening of the simulation. In the bottom panel, the black solid line shows a fit (Eq~\ref{eq:sigma_v}) to the median relation.}\label{fig:resolution}
\end{figure}

In Fig.~\ref{fig:resolution} we show the distribution of $\sigma_x$ and $\sigma_v$ as a function of subhalo mass. When using the peak mass as a proxy of satellite mass, the distributions are quite similar for central and satellites. $\sigma_x$ approaches the softening of the simulation in well resolved subhalos, while it is generally bigger for subhalos with less than $10^{3}$ particles. The velocity scale $\sigma_v$, however, increases with subhalo mass, since more massive objects are dynamically hotter. The median relation can be well fitted by 
\begin{equation}
 \sigma_v=\left(-3.2+7.4\ln N \right) {{\rm km}\, {\rm s}^{-1}}, \label{eq:sigma_v}
\end{equation} where $N$ is the number of particles in the subhalo. However, we caution that the above fitting formula may not be applicable to simulations with different resolutions.

\begin{figure}
 \myplot{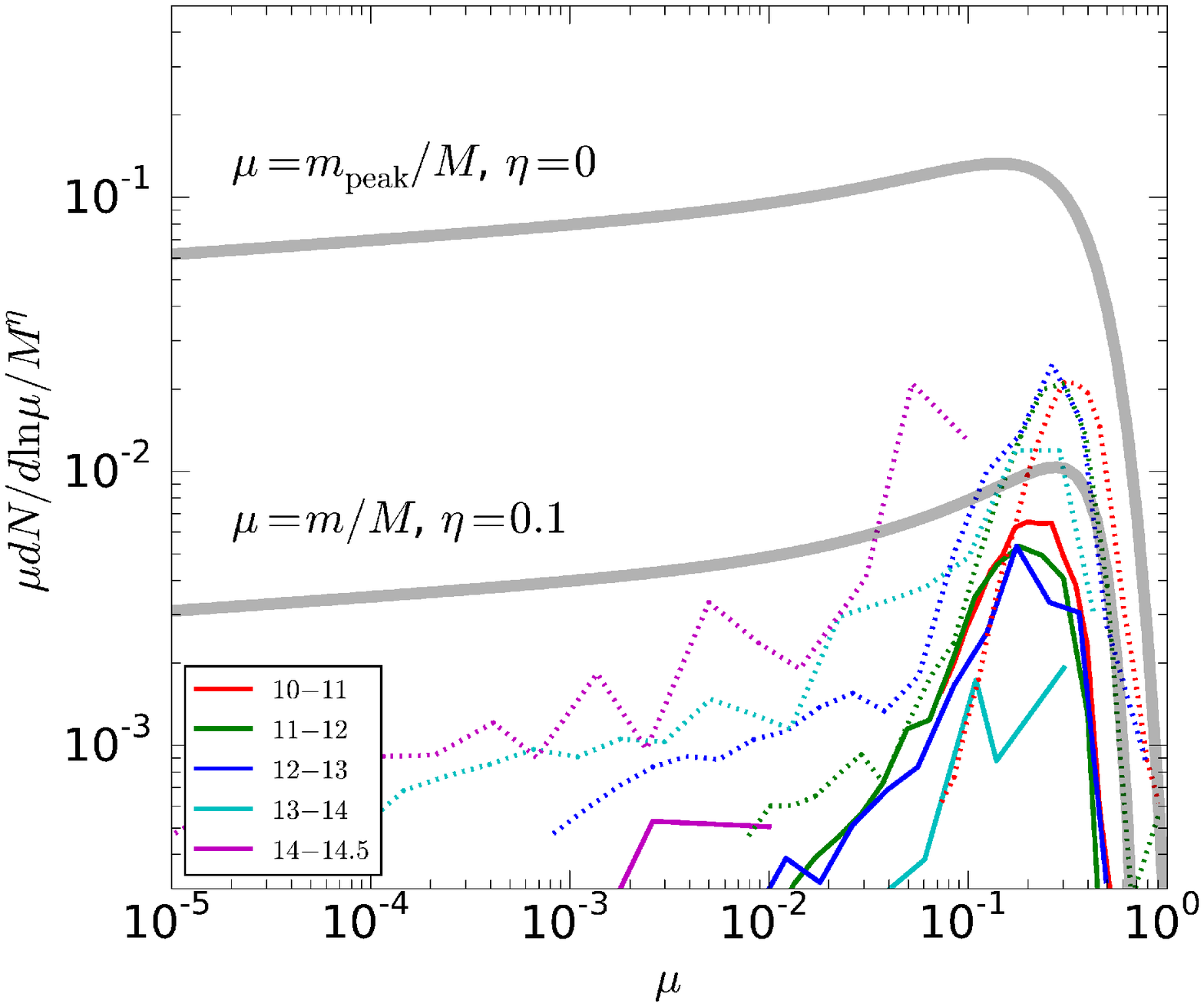}
 \caption{The mass distribution of trapped subhalos that still survive at $z=0$ in Millennium-II. The thick grey lines are the same as in Fig.~\ref{fig:Mill2MF} showing the fitted peak and final mass functions for normal subhalos. The coloured lines show the mass functions of trapped subhalos in different host halo mass bins (as labelled by $\log[M/\msunh]$), with the solid and dotted lines showing the final and peak mass functions respectively.}\label{fig:TrappedMF}
\end{figure}

The mass functions of trapped subhalos are shown in Fig.~\ref{fig:TrappedMF}. These objects are mostly massive objects with $m/M>0.1$, and they evolve only mildly since infall, with the peak and final mass functions differing by a factor of $2 \sim 3$. Note that these trapped subhalos have already been removed in the figures in the main text of the paper. One might wonder whether the over-abundance of subhalos at the high mass end in our result is contaminated by trapped subhalos that are not completely removed. Since the trapped subhalos are mostly located within $2\sigma_x$, they do not contaminate the radial profiles in Fig.~\ref{fig:radialprof} and Fig.~\ref{fig:subfind} in $10^{13}\,\msunh$ halos (corresponding to more than $10^{6}$ particles) where the softening is around $0.002\, R_{200}$.  Subsequently, the existence of the flattening in the subhalo mass function is robust against contaminations from trapped subhalos as discussed in Section~\ref{sec:radial}.

These merged subhalos represent the case when numerical resolution of the simulation is no longer able to separate the subhalo from its host. We have carried out a test using a lower resolution simulation and found that the mass functions of the merged subhalos seem to have converged (though noisily). Even so, whether the galaxies in the trapped subhalos have merged with their central galaxy is a different problem. 

\end{document}